\documentstyle[12pt,epsf]{article}
\newcommand{\beq}{\begin{equation}}
\newcommand{\eeq}{\end{equation}}
\newcommand{\beqa}{\begin{eqnarray}}
\newcommand{\eeqa}{\end{eqnarray}}
\def\simgt{\rlap{\lower 3.5 pt \hbox{$\mathchar \sim$}} \raise 1pt \hbox {$>$}}
\def\simlt{\rlap{\lower 3.5 pt \hbox{$\mathchar \sim$}} \raise 1pt \hbox {$<$}}
\newcommand{\PLB}[3]{Phys.\ Lett.\ {\bf B{#1}}, {#2} ({#3})}
\newcommand{\NPB}[3]{Nucl.\ Phys.\ {\bf B{#1}}, {#2} ({#3})}
\newcommand{\PRD}[3]{Phys.\ Rev.\ {\bf D{#1}}, {#2} ({#3})}
\newcommand{\PRL}[3]{Phys.\ Rev.\ Lett.\ {\bf {#1}}, {#2} ({#3})}
\newcommand{\ZPC}[3]{Z.\ Phys.\ {\bf C{#1}}, {#2} ({#3})}
\newcommand{\etal}{{\em et~al.}}
\newcommand{\sh}{\hat{s}}
\newcommand{\th}{\hat{t}}
\newcommand{\uh}{\hat{u}}

\catcode`\@=11
\def\@citex[#1]#2{\if@filesw\immediate\write\@auxout{\string\citation{#2}}\fi
  \def\@citea{}\@cite{\@for\@citeb:=#2\do
    {\@citea\def\@citea{,\penalty\@m}\@ifundefined
       {b@\@citeb}{{\bf ?}\@warning
       {Citation `\@citeb' on page \thepage \space undefined}}%
\hbox{\csname b@\@citeb\endcsname}}}{#1}}
\def\citer{\@ifnextchar [{\@tempswatrue\@citexr}{\@tempswafalse\@citexr[]}}
\def\@citexr[#1]#2{\if@filesw\immediate\write\@auxout{\string\citation{#2}}\fi
  \def\@citea{}\@cite{\@for\@citeb:=#2\do
    {\@citea\def\@citea{--\penalty\@m}\@ifundefined
       {b@\@citeb}{{\bf ?}\@warning
       {Citation `\@citeb' on page \thepage \space undefined}}%
\hbox{\csname b@\@citeb\endcsname}}}{#1}}
\begin{document}

\begin{titlepage}
\begin{flushright}
        CERN-TH/97-235\\
        RAL-TR-97-044 \\
        NORDITA-97/62 P \\
        TUM/T39-97-21\\
        hep-ph/9709376 \\ 
\end{flushright}
\vskip .8cm
\begin{center}
 \boldmath
{\Large\bf Inelastic photoproduction of polarised $J/\psi$} \unboldmath
%---------------------------------------------------------
\vskip .7cm
{\sc M. Beneke}
\vskip .3cm
{\it Theory Division, CERN, CH-1211 Geneva 23}
%---------------------------------------------------------
\vskip .7cm
{\sc M. Kr\"amer}
\vskip .3cm
{\em Rutherford Appleton Laboratory \\
Chilton, Didcot, OX11 0QX, England}
%---------------------------------------------------------
\vskip .7cm
{\sc M. V\"anttinen
\footnote{\noindent Present address: Institut f\"ur Theoretische Physik, 
Physik-Department der Technischen Universit\"at M\"unchen, 
85747 Garching, Germany}
}
\vskip .3cm
{\em NORDITA, Blegdamsvej 17, DK-2100 Copenhagen \O}
%---------------------------------------------------------
\vskip 1.0cm
\end{center}

\begin{abstract}
\noindent The comparison of $J/\psi$ photoproduction data in the 
inelastic region with theoretical predictions based on the NRQCD
approach has remained somewhat ambiguous and controversial, in
particular at large values of the inelasticity variable $z$.  We study
the polar and azimuthal decay angular distribution of $J/\psi$ mesons
as functions of $z$ and transverse momentum $p_t$. Future measurements
of decay angular distributions at the HERA $ep$ collider will provide a new
test of theoretical approaches to factorisation between perturbation
theory and quarkonium bound-state dynamics and shed light on the
colour-octet production fraction in various regions of $z$ and $p_t$.
\end{abstract}

\vfill

\end{titlepage}

\section{Introduction}
%%%%%%%%%%%%%%%%%%%%%%%%%%%%%%%%%%%%%%%%%%%%%%%%%%%%%%%%%%%%%%%%%%%%%%%
The production of quarkonium in various processes, especially at
high-energy colliders (for reviews, see \cite{BFY,rev}), has been the
subject of considerable interest during the past few years. New data
have been taken at $p\bar{p}$, $ep$ and $e^+e^-$ colliders, and a
wealth of fixed-target data also exist. In theory, progress on
factorisation between perturbative and the quarkonium bound state
dynamics has been made.  The earlier `colour-singlet model' has been
superseded by a consistent and rigorous approach, based on
non-relativistic QCD (NRQCD) \cite{NRQCD}, an effective field theory
that includes the so-called colour-octet mechanisms. On the other
hand, the `colour evaporation' model of the early days of quarkonium
physics \cite{evaporation1} 
has been revived \cite{evaporation}.  Despite these
developments the range of applicability of these approaches to the
practical case of charmonium is still subject to debate, as is the
quantitative verification of factorisation. The problematic aspect is
that, because the charmonium mass is still not very large with respect to
the QCD scale, non-factorisable corrections may not be suppressed
enough, if the quarkonium is not part of an isolated jet, and 
the expansions in NRQCD may not converge very well. In this
situation cross checks between various processes, and predictions of
observables such as quarkonium polarisation and differential cross
sections, are crucial in order to assess the importance of different
quarkonium production mechanisms, as well as the limitations of a
particular theoretical approach.  In this paper we discuss how polar
and azimuthal decay angular distributions of $J/\psi$, produced by real
photons colliding on a proton target in the inelastic region
$p_t>1\,$GeV (or, more conventionally, \mbox{$z \equiv p_\psi 
\cdot p_p / p_\gamma \cdot p_p < 0.9$}), may serve this purpose.

In the NRQCD approach, to which we adhere in this paper, the cross
section for producing a charmonium state $H$ in a photon--proton
collision is written as a sum of factorisable terms,
\beq
  d\sigma(H) = \sum_{i,j\in\{q,g,\gamma\}} \int dx_1 dx_2\,
  f_{i/\gamma}(x_1) f_{j/p}(x_2)\,\sum_{n} 
  d\hat{\sigma}(i+j\to c\bar c[n])\,
               \langle {\cal O}_n^H\rangle ,
  \label{NRQCDsum}
\eeq
where $n$ denotes the colour, spin and angular momentum state of an
intermediate $c\bar c$ pair and $f_{i/\gamma}$ and $f_{j/p}$ the
parton distributions in the photon and the proton, respectively. The
short-distance cross sections $d\hat{\sigma}(i+j\to c\bar c[n])$ can
be calculated perturbatively in the strong coupling $\alpha_s$.  The
matrix elements $\langle {\cal O}_n^H \rangle \equiv \langle 0|{\cal
  O}_n^H|0 \rangle$ (see \cite{NRQCD} for their definition) are
related to the non-perturbative transition probabilities from the
$c\bar{c}$ state $n$ into the quarkonium $H$.  The magnitude of these
probabilities is determined by the intrinsic velocity $v$ of the bound
state. Thus the above sum is a double expansion in $\alpha_s$ and $v$.

Within NRQCD the leading term in $v$ to inelastic photoproduction of
$J/\psi$ comes from an intermediate $c\bar{c}$ pair in a
colour-singlet $^3\!S_1$ state and coincides with the colour-singlet
model result.  (The notation for the angular momentum configuration is
$^{2S+1}\!L_J$ with $S$, $L$ and $J$ denoting spin, orbital and total
angular momentum, respectively.) Cross sections \cite{BJ}, polar
\cite{CSMpred1}, and polar and azimuthal \cite{CSMpred2} decay angular
distributions have been calculated for the direct-photon contribution,
in which case $i=\gamma$ and $f_{\gamma/\gamma}(x)=\delta(1-x)$ in
(\ref{NRQCDsum}).  The angular integrated cross section is known to
next-to-leading order (NLO) in $\alpha_s$ \cite{Kraemer}. 
The colour-singlet contribution, including next-to-leading corrections
in $\alpha_s$, is known to reproduce the unpolarised data adequately.
But there is still a considerable amount of uncertainty in the
normalisation of the theoretical prediction, which arises from the
value of the charm quark mass and the wave-function at the origin, as
well as the choice of parton distribution functions and
renormalisation/factorisation scale.  

At order $v^4\sim 0.05$--$0.1$ relative to the colour-singlet
contribution, the $J/\psi$ can also be produced through intermediate
colour-octet $^3\!S_1$, $^1\!S_0$ and $^3\!P_J$ configurations. In the
inelastic region, they have been considered in \cite{Cacciari,Ko} for
the direct photon contribution and in \cite{Cacciari2} for resolved 
photons, in which case
the photon participates in the hard scattering through its parton
content. Colour-octet contributions to the total
photoproduction cross section (integrated over all $z$ and $p_t$) are
known to next-to-leading order \cite{MMP97}.  The polarisation of
inelastically produced $J/\psi$ due to these additional production
mechanisms, however, has not been calculated so far. 

Because the colour-octet contributions are suppressed as $v^4$, but,
in the inelastic region, contribute at the same order in $\alpha_s$ as
the colour-singlet contribution, they are of interest only if they are
enhanced by other factors, either numerical or kinematical.  In this
respect the situation is similar to a certain $v^2$-correction, which 
arises already in the colour-singlet model \cite{Jung} and 
becomes kinematically enhanced at $z$ close to 1. The colour-octet
production channels are indeed kinematically different from the
colour-singlet one, because the $^1\!S_0$ and $^3\!P_J$ configurations
can be produced through $t$-channel exchange of a gluon already at
lowest order in $\alpha_s$. (For the $^3\!S_1$ octet this is true for
the resolved process.) This leads to a significantly enhanced
amplitude, in particular in the large-$z$ region. 
The colour-octet contributions to $J/\psi$ photoproduction 
are indeed strongly peaked at large $z$ \cite{Cacciari,Ko}. Such a
shape is not supported by the data, which at first sight could lead to
a rather stringent constraint on the octet matrix elements $\langle
{\cal O}_8[n]\rangle$, $n\in\{^1\!S_0,^3\!P_0\}$ and to an
inconsistency with the values obtained for these matrix elements from
other processes. However, the peaked shape of the $z$-distribution is
derived neglecting the energy transfer in the non-perturbative
transition $c\bar{c}[n]\to J/\psi+X$.  In reality the peak may be
considerably smeared \cite{BRW} as a consequence of resumming
kinematically enhanced higher-order corrections in $v^2$ and no
constraint or inconsistency can be derived from the endpoint behaviour
of the $z$-distribution at present. As a consequence, the role of
octet contributions to the direct process remains unclear.  The
resolved photon contribution, on the other hand, could be entirely
colour-octet dominated \cite{Cacciari2,Kniehl}. The $z$-distribution
should then begin to rise again at small $z$, if the colour-octet
matrix elements are as large as suggested by NRQCD velocity scaling
rules \cite{LMNMH92,NRQCD} and fits to hadroproduction data.

Our motivation for considering the decay angular distributions,
including all direct and resolved production mechanisms, is to provide
another observable that can clarify the relative importance of
colour-octet production in photoproduction in different kinematic
regions. Many of the above-mentioned uncertainties and difficulties do
not affect the polarisation yield. For example, the resummation that is 
necessary in the endpoint region may lead to a significant
redistribution of $d\sigma/dz$ in $z$, but affects the normalised
decay angular distributions to a lesser degree, if they 
do not have a strong $z$-dependence in the region affected by the 
smearing. We find that some angular coefficients, especially those
for the azimuthal angle dependence, take essentially different values
in the colour-singlet and colour-octet processes. A measurement of
decay angular distributions would therefore provide information on the
relevance of colour-octet production, also at $z$ close to 1, which is
largely independent of normalisation uncertainties.

The polarisation of the $J/\psi$ can be determined by measuring the
angular distribution of the leptonic decay $J/\psi \rightarrow
l^+ l^-$. To date, experimental measurements of $J/\psi$
polarisation exist only for diffractive (elastic and proton
dissociation) photoproduction \cite{H1,ZEUSelastic}, to which the
inclusive formalism of NRQCD does not apply, and for fixed-target
hadroproduction \cite{FThadro}. The latter can be compared with
predictions obtained in the colour-singlet model \cite{FTpred1} and
NRQCD \cite{FTpred2,FTpred3} for $p_t$-integrated cross sections. As
discussed in \cite{rev} the experimental finding of no polarisation is
only marginally consistent with the NRQCD prediction. Photoproduction
offers another opportunity to learn about whether the $J/\psi$
polarisation carries information on the spin of the heavy quark pair
produced at short distances, which is expected in theoretical
approaches in which spin symmetry is at work. With the expected
increase in luminosity at the HERA $ep$ collider, polarisation in
photoproduction of $J/\psi$ at different values of $z$ and $p_t$ could
provide an attractive diagnostic tool in addition to the widely
discussed polarisation measurement in $p\bar{p}$ collisions at the
Fermilab Tevatron \citer{CDFpred,AKL97}.

The paper is organised as follows: Section~\ref{mechanisms} discusses
the production mechanisms and calculational details regarding decay
angular distributions. In Section~\ref{theory} we pause for
theoretical considerations that influence our choice of cuts and other
parameters in the analysis. Section~\ref{results} presents results and
their discussion, followed by a summary in Section~\ref{summary}.
Appendix~\ref{pfs} contains the covariant definitions of coordinate
systems and polarisation vectors and Appendix~\ref{dmes} summarises
the density matrices for all subprocesses considered in the paper.

\section{Production mechanisms and cross sections \label{mechanisms}}
%%%%%%%%%%%%%%%%%%%%%%%%%%%%%%%%%%%%%%%%%%%%%%%%%%%%%%%%%%%%%%%%%%%%%%%
We assume that the $J/\psi$ transverse momentum $p_t>1\,$GeV, in order
to suppress the diffractive contribution and higher-twist corrections
in general. Away from $p_t=0$ (or $z=1$) the leading-twist hard subprocesses
contributing to inelastic $J/\psi$ production can be classified as
follows:

\begin{enumerate}

\item Direct photon mechanisms. At leading order in the
      strong coupling constant, $O(\alpha\alpha_s^2)$, these are
\beqa
\gamma + g & \rightarrow & c\bar c 
\left[ ^3\!S_1^{(1)}, \,^3\!S_1^{(8)}, \,^1\!S_0^{(8)}, \,^3\!P_J^{(8)}
 \right] + g, \label{subpg}\\
\gamma + q/\bar{q} & \rightarrow & c\bar c
\left[ ^3\!S_1^{(8)}, \,^1\!S_0^{(8)}, \,^3\!P_J^{(8)} \right] + q/\bar{q},
\eeqa
where the initial-state parton originates from the target proton. 
 
\item Resolved photon mechanisms. At leading order,
      $O(\alpha_s^3)$, the subprocesses are
\beqa
g + g      & \rightarrow & c\bar c
\left[ ^3\!S_1^{(1)}, \,^3\!S_1^{(8)}, \,^1\!S_0^{(8)}, \,^3\!P_J^{(8)}
 \right] + g, \label{subgg} \\
g + q/\bar{q}      & \rightarrow & c\bar c
\left[ ^3\!S_1^{(8)}, \,^1\!S_0^{(8)}, \,^3\!P_J^{(8)} \right] + q/\bar{q}, \\
q + \bar q & \rightarrow & c\bar c
\left[ ^3\!S_1^{(8)}, \,^1\!S_0^{(8)}, \,^3\!P_J^{(8)} \right] + g,
\eeqa
where one of the initial-state partons originates from the photon and
the other originates from the proton.

\end{enumerate}

\noindent The direct-photon mechanisms dominate in the region 
$z \;\simgt\; 0.2$, whereas resolved-photon mechanisms become
important in the region $z \;\simlt\; 0.2$. (These numbers depend on
the values of the colour-octet matrix elements, as well as on the
$p_t$-cut.) At HERA energies, photon--quark fusion can contribute about
10\%--15\% to the cross section at large $z$. Quark--gluon fusion
constitutes about 20\%--40\% of the resolved cross section at $z
\;\simlt\; 0.2$ and becomes more important than gluon--gluon fusion at
larger $z$.  Quark--antiquark fusion is always completely negligible.

The above list includes those colour-octet production channels that
are suppressed by at most $v^4$ relative to the leading colour singlet
production channel. The suppression of the octet contributions follows
from a multipole expansion of the non-perturbative transition
$c\bar{c}[n]\to J/\psi+X$.  From a $^3\!P_J^{(8)}$ intermediate state,
the physical $J/\psi$ state can be reached by a single chromoelectric
dipole transition, from a $^3\!S_1^{(8)}$ state by two consecutive
electric dipole transitions, and from a $^1\!S_0^{(8)}$ state by a
chromomagnetic dipole transition. Each electric dipole transition
brings a factor $v^2$, and the magnetic dipole transition a factor
$v^4$. In addition, the hard production vertex for a $P$-wave $c\bar
c$ state is suppressed already by $v^2$ relative to production in an
$S$-wave state. In photon-gluon fusion, the $^3\!S_1^{(8)}$-amplitude
is kinematically identical to the $^3\!S_1^{(1)}$-amplitude. The
$^3\!S_1^{(8)}$-channel is therefore insignificant for the direct-photon 
contribution.

In resolved photon interactions, on the other hand, the
$^3\!S_1^{(8)}$-channel dominates at $p_t\;\simgt\; 5\,$GeV, 
because it includes
a gluon fragmentation component \cite{BF}, in both the gluon--gluon and
gluon--quark fusion contributions, and therefore falls only as $1/p_t^4$
at large $p_t$. The resolved photon amplitudes are identical to those
relevant to $J/\psi$ production in hadron--hadron collisions
\cite{ChoL,BEN97} and at HERA energies the relative importance of the
various contributions as functions of $p_t$ is nearly the same as at
Tevatron energies.

The direct-photon mechanisms above all decrease at least as $1/p_t^6$
at large $p_t$, with the exception of $\gamma+q\to
c\bar{c}[^3\!S_1^{(8)}]+q$. Fragmentation contributions in
photon-gluon fusion exist only at the next order in $\alpha_s$.  They
exceed the leading-order contributions at $p_t\;\simgt\; 10\,$GeV
\cite{Godbole,Kniehl}. We therefore conclude that our list 
includes all important leading-twist production mechanisms for all $z$
and as long as $p_t\;\simlt\; 10\,$GeV.

We expect that higher-twist corrections due to multiple interactions
with the proton or photon remnant would be suppressed as a power of 
$\Lambda^2/(Q^2+p_t^2)$, where $\Lambda\;\simlt\;1\,$GeV is a typical
QCD scale and $Q$ is one of the scales involved in the bound state
dynamics, $Q \approx m_c, m_c v$, or $m_c v^2$. Since $m v^2 \sim
\Lambda$ for charmonium and bottomonium, one may expect large
higher-twist corrections at small $p_t$, when the heavy
quark--antiquark pair moves parallel with a remnant jet and remains in
its hadronization region over a time $1/\Lambda$ in the quarkonium
rest frame.\footnote{Some aspects of higher-twist corrections to
  $\gamma+g\to c\bar{c}[^3\!S_1^{(1)}]+g$ have been considered in
  \cite{Ma}, with the surprising conclusion that the higher-twist
  correction is $\Lambda^2/(4 m_c^2)$, even at very large $p_t$,
  rather than $\Lambda^2/p_t^2$. The term that does not scale as
  $\Lambda^2/p_t^2$ enters in the combination $e_1(z)+4 d_1(z)$, where
  $e_1$ and $d_1$ are certain twist-4 multi-parton correlation
  functions defined in \cite{Ma}.  However, in the approximation
  considered in \cite{Ma} one finds $e_1=4 d_1$. If there existed a
  sign inconsistency in \cite{Ma}, the $\Lambda^2/(4 m_c^2)$-term
  would disappear and the result conform to our intuition.}

The differential cross section for $J/\psi$ production and its
subsequent leptonic decay $J/\psi\to l^+l^-$ through any of the
resolved-photon subprocesses can be written as
\beqa
  \frac{1}{B_{ll}}\,\frac{d\sigma^{i j}}{d\Omega dz dp_t}
  &=& \int\limits_{x_{1,min}}^1 \!dx_1\,f_{i/\gamma}(x_1,\mu_F) 
       f_{j/p}(x_2,\mu_F)\,\frac{1}{16\pi\hat{s}^2}\,\frac{2 x_1 x_2 p_t }
       {z (x_1-z)} \\
  & & \hspace*{-1.8cm}\times\frac{3}{8\pi} \left[ \, 
      \rho^{ij}_{11} + \rho^{ij}_{00} 
        + (\rho^{ij}_{11} - \rho^{ij}_{00}) \cos^2\theta 
        + \sqrt{2} \, {\rm Re} (\rho^{ij}_{10}) \sin 2\theta \cos\phi
        + \rho^{ij}_{1,-1} \sin^2\theta \cos 2\phi \, \right] \nonumber ,
\eeqa
where $B_{ll}$ is the $J/\psi\to l^+l^-$ branching ratio, $s =
(p_\gamma + p_p)^2$, $\hat{s}=x_1 x_2 s$ and the parton distribution
of the proton is evaluated at
\beq
  x_2 = \frac{x_1 p_t^2 + M^2 (x_1-z)}{sz (x_1-z)}
\eeq
with factorisation scale $\mu_F$. Here and in the following we use
$M=2 m_c$. The variables $z$ and $p_t$ are subject to the restriction
\begin{equation}
(1-z) (s z-M^2) > p_t^2
\end{equation}
and
\begin{equation}
x_1 > x_{1,min} = \frac{z (sz-M^2)}{s z-p_t^2-M^2}.
\end{equation}
The angles $\theta$ and $\phi$ refer to the polar and azimuthal angle
of the $l^+$ in the $J/\psi$ decay with respect to a coordinate system
defined in the $J/\psi$ rest frame. (See Appendix~\ref{pfs} for
details on their definition.) Finally
\beq
  \rho^{ij}_{\lambda\lambda'} \equiv 
  A[ij \rightarrow J/\psi(\lambda) + X]\,
  A^*[ij\rightarrow J/\psi(\lambda') + X] 
  \label{rhodef}
\eeq
are density matrix elements for $J/\psi$ production, where a summation
(average) over the spins of $X$ ($i,j$) is understood. The kinematical
relations for the direct-photon process follow from setting $i=\gamma$
and $f_{\gamma/\gamma}(x_1,\mu_F)=\delta(1-x_1)$.

The polarisation analysis in NRQCD \cite{BR96,FTpred3,polarisation} is
based on the symmetries of the NRQCD Lagrangian, of which spin and
rotational symmetry are crucial. In electric dipole transitions the
heavy quark spins remain intact, so that the $J/\psi$ spin orientation
will be the same as the perturbatively calculable orientation of the
total $c\bar{c}$ spin ${\bf s}_q + {\bf s}_{\bar q}$ in the
intermediate state. The $^1\!S_0^{(8)}$ intermediate state is
rotationally invariant and leads to random orientation of the $J/\psi$
spin. Technically, we have
\beq
\label{sep}
 \rho^{ij}_{\lambda\lambda'} =  
 \rho^{ij}_{\lambda\lambda'}[^3\!S_1^{(1)}] + 
 \rho^{ij}_{\lambda\lambda'}[^3\!S_1^{(8)}] +
 \rho^{ij}_{\lambda\lambda'}[^1\!S_0^{(8)}] +
 \rho^{ij}_{\lambda\lambda'}[\{S=1,L=1\}^{(8)}] + \ldots,
\eeq
where $\rho^{ij}_{\lambda\lambda'}[n]$ refers to production through a
$c\bar{c}$ pair in a state $n$.  The above decomposition implies that
no interference occurs between the amplitudes for the different terms
in the sum. The symmetries of NRQCD do not forbid interference of
different $^3\!P_J$-states. One finds \cite{BR96}
\begin{eqnarray} 
\rho^{ij}_{\lambda\lambda'}[\{S=1,L=1\}^{(8)}] 
&\propto& \sum_{L_z} A[ij\to c\bar{c}[(1 L_z;1\lambda)]+X]\,
A^*[ij\to c\bar{c}[(1 L_z;1\lambda')]+X] \nonumber\\
&&\hspace*{-3cm}
\not=\,\sum_{J=0,1,2}\rho^{ij}_{\lambda\lambda'}[^3\!P_J^{(8)}],
\end{eqnarray}
where the quantum numbers of the $c\bar{c}$ pair refer to $(L L_z,S
S_z)$.  NRQCD factorisation implies that the density matrices can be
written as
\beq
\rho^{ij}_{\lambda\lambda'}[n]=K^{ij}[n]_{ab\ldots}\,
\langle {\cal O}^{J/\psi}_{\lambda\lambda'}[n]_{ab\ldots}\rangle,
\eeq
where $\langle {\cal
  O}^{J/\psi}_{\lambda\lambda'}[n]_{ab\ldots}\rangle$ is a NRQCD
matrix element with Cartesian indices $a,b,\ldots$, and
$K^{ij}[n]_{ab\ldots}$ the corresponding short-distance coefficient.
The final step is a tensor decomposition of these matrix elements,
which, in the case of interest, can be formulated as a projection of
the $c\bar{c}$ production amplitude. For $J/\psi$ production at the
considered order in $v^2$, the symmetries of NRQCD are sufficient to
reduce all non-perturbative input to the four parameters $\langle
{\cal O}^{J/\psi}[n]\rangle$ with
$n\in\{^3\!S_1^{(1)},^3\!S_1^{(8)},^1\!S_0^{(8)},^3\!P_0^{(8)}\}$
defined as for {\em unpolarised} $J/\psi$ production.

The calculation then consists of evaluating the density matrix
elements for each separate term in (\ref{sep}) and all partonic
subprocesses. We express these matrices as
\beqa
  \rho_{\lambda\lambda'}^{ij}[n]
  &=& A^{ij}[n] \, [\epsilon^*(\lambda)\cdot\epsilon(\lambda')]
        \nonumber \\
  & & \mbox{} + M^2 B^{ij}[n]\,[ p_1\cdot\epsilon^*(\lambda) \; 
  p_1\cdot\epsilon(\lambda')] + M^2 C^{ij}[n]\,[ p_2\cdot\epsilon^*(\lambda) 
  \; p_2\cdot\epsilon(\lambda')]
        \nonumber \\
  & & \mbox{} + M^2 D^{ij}[n]\,[ p_1\cdot\epsilon^*(\lambda) \; 
  p_2\cdot\epsilon(\lambda')
  + p_2\cdot\epsilon^*(\lambda) \; p_1\cdot\epsilon(\lambda')] ,
  \label{rhogeneral}
\eeqa
where $\epsilon(\lambda)$ is the $J/\psi$ polarisation vector, $p_1$
is the momentum of the photon (or the parton originating from the
photon in resolved contributions), and $p_2$ is the momentum of the
parton in the target. The coefficients $A,B,C,D$ are independent of
the choice of axes in the $J/\psi$ rest frame and proportional to a
NRQCD matrix element. Their analytic expressions are collected in
Appendix \ref{dmes}.

The decay angular distribution in the $J/\psi$ rest
frame is often parametrised as
\vspace*{0.2cm}
\beq
  \frac{d\sigma}{d\Omega dy}
  \propto  
  1 + \lambda(y) \cos^2\theta + \mu(y) \sin 2\theta \cos\phi
  + \frac{\nu(y)}{2} \sin^2\theta \cos 2\phi,
  \label{paramdef}
\vspace*{0.2cm}
\eeq
where $y$ stands for a set of variables and 
$\lambda, \mu, \nu$ are obviously related to (appropriate integrals 
of) 
the density matrix elements as
\vspace*{0.2cm}
\beq
  \lambda =\frac{\rho_{11} - \rho_{00}}{\rho_{11} + \rho_{00}}, \qquad
  \mu     =\frac{\sqrt{2}\,{\rm Re}\,\rho_{10}}{\rho_{11} + \rho_{00}}, 
  \qquad
  \nu     =\frac{2 \rho_{1,-1}}{\rho_{11} + \rho_{00}}.
  \label{relation}
\vspace*{0.2cm}
\eeq
Because of the dependence of $\epsilon(\lambda)$ on the 
definition of a coordinate system (see Appendix~\ref{pfs}), the parameters
$\lambda,\mu,\nu$ depend on this definition. 

\section{Theoretical considerations \label{theory}} 
%%%%%%%%%%%%%%%%%%%%%%%%%%%%%%%%%%%%%%%%%%%%%%%%%%%%%%%%%%%%%%%%%%%%%%
In this Section we discuss some theoretical issues that influence our
choice of cuts. We also motivate the values of NRQCD long-distance
matrix elements that we subsequently use.

The NRQCD expansion of the quarkonium production cross section applies
to the leading-twist contribution of an inclusive production cross
section. Leading-twist means that the result is accurate up to
corrections that scale as some power of $\Lambda/m_c$ in the limit
that $m_c\to\infty$.  Up to such corrections, NRQCD also applies to the
total $J/\psi$ photoproduction cross section. The leading contribution
is $O(\alpha\alpha_s)$ and purely colour-octet
\cite{Cacciari,Amundson}.  It formally contributes only at $z=1$,
$p_t=0$, i.e. in the diffractive region. Soft-gluon emission
during conversion of the colour-octet $c\bar{c}$ pair into a $J/\psi$
is expected to `smear' the delta-functions at $z=1$ and $p_t=0$ over a
region $\delta z\sim 0.25$, $\delta p_t\sim 0.5\,$ GeV \cite{BRW}.
One may ask whether the experimentally measured diffractive $J/\psi$ cross
section (with or without proton dissociation) could be considered as
part of the leading-twist total cross section. Or whether it should be
considered as a pure higher-twist phenomenon which cannot be regarded
as dual (in the sense of parton--hadron duality) to the
$O(\alpha\alpha_s)$ contribution in the inclusive formalism.

In order for the first possibility to be realised, the soft gluons, 
which are emitted in the transition of the colour-octet $c\bar{c}$
pair into $J/\psi+X$, would have to recombine into a proton or a
low-mass diffractive final state. Although it cannot be argued from
first principles against this possibility, it certainly appears
unlikely.  It would also hardly be compatible with the factorisation
assumption of NRQCD that the above colour neutralisation is universal,
i.e.  independent of the rest of the process, again up to higher-twist
corrections. (Clearly, complete independence is not possible, because
some colour exchange between $J/\psi+X$ and the rest of the process is
necessary, if the $J/\Psi+X$ state originates from a colour-octet
$c\bar{c}$-pair.)

The clearest indication that the diffractive contribution should be
considered as a higher-twist correction, which is not part of a
leading-twist calculation of NRQCD, is experimental. The H1
collaboration has measured \cite{H1} the polar decay angle
distribution and the ZEUS collaboration has measured
\cite{ZEUSelastic} the polar and azimuthal decay angular distribution.
Models of diffractive production based on hard two-gluon
\cite{Ryskin1,Brodsky} or soft-pomeron \cite{Donnachie} exchange
predict $\lambda=1$ \cite{Ryskin1} ($\lambda$ is defined in
(\ref{paramdef})), in agreement with the HERA measurements and earlier
fixed-target data \cite{FTelastic}. On the other hand, the
polarisation signature of the leading-twist parton reaction $\gamma g
\rightarrow c\bar c$ is identical to the signature in the process $gg
\rightarrow c\bar c$ \cite{FTpred3}.  The result is $\lambda=0$ if the
$^1\!S_0$ configuration dominates and $\lambda=1/2$ if $^3\!P_J$
dominates. Any linear combination of these values is incompatible with
the experimental data.  Since the diffractive cross section (according
to the experimental definitions of \cite{H1,ZEUSelastic}) is about as
large as the inelastic cross section \cite{H1,ZEUSinelastic}, we
conclude that NRQCD cannot be used to predict the photoproduction
cross section integrated over all $z$ and $p_t$.

In order to apply NRQCD we therefore have to cut the elastic region
without restricting the inclusive nature of the process. The HERA
collaborations conventionally define the inelastic region through the
requirement $z<0.9$. Let us now argue that it is theoretically
advantageous to define the inelastic region through a cut in $p_t$.
It is obvious theoretically, and confirmed experimentally, that the
slope of the $p_t$-distribution is significantly smaller for inelastic
production than for elastic production (with or without proton
dissociation). A $p_t$-cut at $p_t > 1\,$GeV already eliminates most of
the diffractive contribution as well as higher-twist corrections in
general and no further cut on $z$ is necessary. In fact, the cross
section with an additional cut $z<0.9$ {\em cannot} be reliably
predicted in NRQCD. As emphasised in \cite{BRW}, because the NRQCD
expansion is singular at $z=1$, only an average cross section over a
sufficiently large region close to $z=1$ can be predicted. The
$z$-distribution itself requires additional non-perturbative
information in the form of so-called shape functions. These
shape functions are also required to predict the $p_t$-distribution
with an additional cut $z<0.9$, but not if $z$ is
integrated up to its kinematic maximum. In the following, we define
the inelastic region through the cut $p_t > 1\,$GeV. If statistics is
not a limitation, it might be preferable to use $p_t>2\,$GeV to
further suppress the higher-twist contributions and difficulties in
predicting the $p_t$-distribution at low $p_t$, because of
(perturbative) soft-gluon emission. Note that the resummation of 
higher-order $v^2$-corrections in NRQCD will also cause some smearing in 
transverse momentum, which we expect to be less important than that 
caused by perturbative soft-gluon emission. 

Because the colour-octet contributions to inelastic $J/\psi$
production are strongly enhanced at large $z$, an immediate
consequence of integrating up to $z_{max}$ rather than 0.9 is that the
$p_t$-distribution is now dominated by colour-octet production, as
will be discussed in more detail below. The suggested importance of
the colour-octet mechanisms could be further investigated 
experimentally, if hadronic activity in the vicinity of the $J/\psi$
could be detected. If a $J/\psi$ is produced through a colour-octet
$c\bar{c}$ pair, we expect it to be accompanied by light hadrons more often 
than if it is produced through a colour-singlet pair.

The cross sections and decay angular distributions depend on four
parameters related to the probability of the transition
$c\bar{c}[n]\to J/\psi+X$. The colour-singlet matrix element can be
related to the $J/\psi$ wave-function at the origin. For $\langle
{\cal O}_8^{J/\psi}({}^3\!S_1)\rangle$ we use the value obtained 
in \cite{BEN97} from a fit to
hadroproduction of $J/\psi$ at large $p_t$. Its precise
numerical value does not influence our analysis, because the
$^3\!S_1$-colour-octet channel is important only for resolved photons
at large transverse momentum. Our predictions do depend crucially on
$\langle {\cal O}_8^{J/\psi}({}^1\!S_0)\rangle$ and $\langle {\cal
  O}_8^{J/\psi}({}^3\!P_0)\rangle$, both of which are not very well
known. The following constraints can be obtained from other $J/\psi$
production processes,
\begin{eqnarray}
\langle {\cal O}_8^{J/\psi}({}^1\!S_0)\rangle+\frac{3.5}{m_c^2}
\langle {\cal O}_8^{J/\psi}({}^3\!P_0)\rangle
&=& (3.90\pm 1.14)\cdot 10^{-2}\,\mbox{GeV}^3
\qquad\mbox{(Tevatron \cite{BEN97})} \nonumber \\
\langle {\cal O}_8^{J/\psi}({}^1\!S_0)\rangle+\frac{7}{m_c^2}
\langle {\cal O}_8^{J/\psi}({}^3\!P_0)\rangle
&=& 3.0 \cdot 10^{-2}\,\mbox{GeV}^3
\quad\mbox{(fixed-target hadroproduction \cite{FTpred3})} \nonumber \\
\langle {\cal O}_8^{J/\psi}({}^1\!S_0)\rangle+\frac{3.6}{m_c^2}
\langle {\cal O}_8^{J/\psi}({}^3\!P_0)\rangle
&<& 2.8 \cdot 10^{-2}\,\mbox{GeV}^3
\qquad\mbox{($B\to J/\psi X$)} \nonumber 
\end{eqnarray}
where $m_c=1.5\,$GeV is assumed. For various reasons, all of these
determinations should probably be considered as uncertain within a
factor of 2. The constraint from inclusive $B$ decays has been
obtained from the leading-order calculation of \cite{Ko,FHMN}, setting the
colour-singlet contribution, whose magnitude is rather uncertain, to
zero. (With the parameters of \cite{FHMN}, we would have obtained 
$4.4 \cdot 10^{-2}\,\mbox{GeV}^3$ instead of $2.8 \cdot 10^{-2}\,
\mbox{GeV}^3$.) Including the colour-singlet contribution would strengthen the
inequality considerably, but this cannot be justified given  
the NLO result of \cite{BE}. In view of these uncertainties and given that
they do not allow us to constrain separately 
$\langle {\cal O}_8^{J/\psi}({}^1\!S_0)\rangle$ and 
$\langle {\cal O}_8^{J/\psi}({}^3\!P_0)\rangle$ with confidence, 
we consider two
scenarios in which the constraints are (approximately) saturated
either by $\langle {\cal O}_8^{J/\psi}({}^1\!S_0)\rangle$ or $\langle
{\cal O}_8^{J/\psi}({}^3\!P_0)\rangle$ alone.

The values of all parameters are summarised in Table~\ref{tab1}.
Further constraints could be obtained from the $p_t$-distribution in
photoproduction, if all kinematically allowed $z$ are integrated over.
However, for the reasons mentioned earlier, no constraint can be
derived from the endpoint region of the $z$-distribution.

\begin{table}[t]
\addtolength{\arraycolsep}{0.1cm}
\renewcommand{\arraystretch}{1.25}
$$
\begin{array}{c|cccc}
\hline\hline
\mbox{Scenario} & \langle {\cal O}_1^{J/\psi}({}^3\!S_1)\rangle & 
   \langle {\cal O}_8^{J/\psi}({}^3\!S_1)\rangle & 
   \langle {\cal O}_8^{J/\psi}({}^1\!S_0)\rangle &
   \langle {\cal O}_8^{J/\psi}({}^3\!P_0)\rangle/m_c^2 \\ 
\hline
\mbox{I}  & 116 & 1.06 & 3.0 & 0.0 \\
\mbox{II} & 116 & 1.06 & 0.0 & 1.0 \\
\hline\hline
\end{array}
$$
\caption{\label{tab1}
Values of the NRQCD matrix elements in $10^{-2}\,$GeV$^3$ taken for 
the analysis; $m_c=1.5\,$GeV.}
\end{table}

\section{Results  \label{results}}
%%%%%%%%%%%%%%%%%%%%%%%%%%%%%%%%%%%%%%%%%%%%%%%%%%%%%%%%%%%%%%%%%%%%%%
\subsection{Cross sections} 

We begin with differential cross sections in order to display the 
relative magnitude of the various contributions, whose different 
polarisation yield will influence the decay angular distributions.

\begin{figure}[p]
   \vspace{-2.3cm}
   \epsfysize=16.8cm
   \epsfxsize=12cm
   \centerline{\epsffile{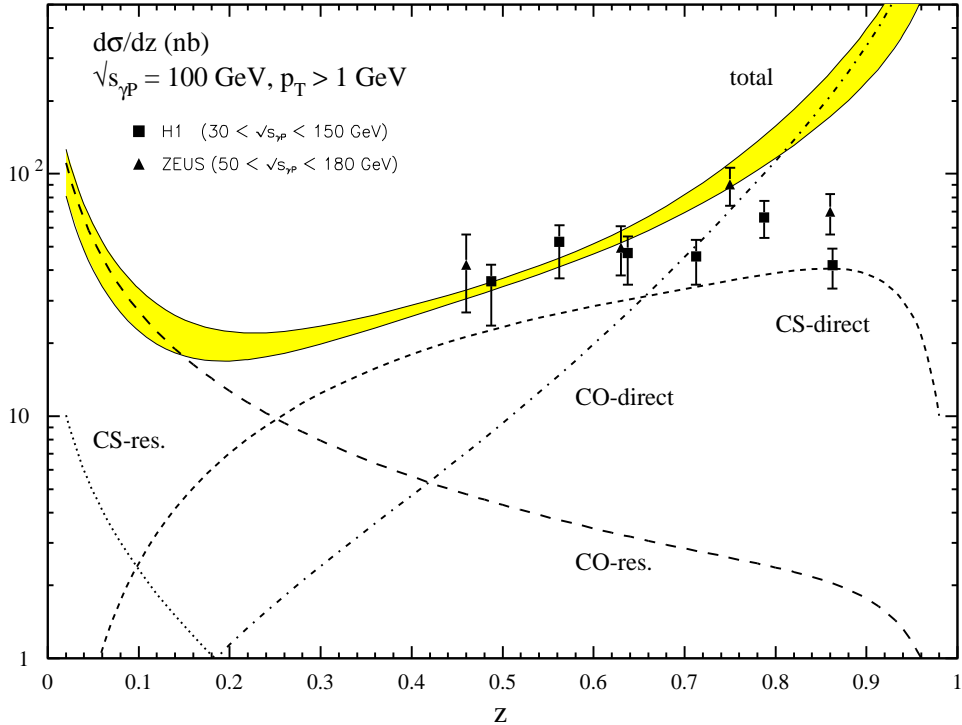}}
   \vspace*{-4cm}
\caption[dummy]{\small \label{sigmaz}
  Colour-singlet (CS) and colour-octet (CO) contributions due to
  direct and resolved photons to the $J/\psi$ energy distribution
  $d\sigma/dz$ at the photon--proton centre-of-mass energy
  $\sqrt{s_{\gamma p}} =100\,$GeV in comparison with HERA data
  \cite{H1,ZEUSinelastic} averaged over the specified range of
  $\sqrt{s_{\gamma p}}$. The shaded area bounded by the solid lines
  represents the sum of all contributions according to scenarios I and
  II for the colour-octet matrix elements. The lines corresponding to
  separate colour-octet contributions are plotted for $\langle {\cal
    O}_8^{J/\psi}(^1\!S_0)\rangle= \langle {\cal
    O}_8^{J/\psi}(^3\!P_0)\rangle/m_c^2=0.008\,\mbox{GeV}^3$. The
  colour-singlet cross section is evaluated in leading order in
  $\alpha_s$.  Other parameters: $m_c=1.5\,$GeV,
  renormalisation/factorisation scale $\mu=2 m_c$, GRV LO proton and
  photon parton distributions \cite{GRV},
  $\Lambda^{(4)}_{LO}=200\,$MeV.}
\end{figure}

The $J/\psi$ energy distribution is shown in Figure~\ref{sigmaz} as a 
function of the scaling variable $z=p_\psi \cdot p_p / p_\gamma \cdot
p_p$ for a typical HERA photon--proton centre-of-mass energy
$\sqrt{s_{\gamma p}}=100\,$GeV and compared with H1 \cite{H1} and ZEUS
\cite{ZEUSinelastic} data.  (Apart from slightly different
colour-octet matrix elements, the presentation coincides with that of
\cite{Cacciari2}.) The colour-octet contributions exceed the
colour-singlet contribution both for large $z \;\simgt\;0.65$ and
for small $z \;\simlt\; 0.25$.

%%%%%%%%%%%%%%%%%%%%%%%%%%%%%%%%%%%%%%%%%%%%%%%%%%%%%%%%%%%%%%%%%%%%%%%%%%
%%%%%%%%%%%%%%%%%%%%%%%%%%%%%%%%%%%%%%%%%%%%%%%%%%%%%%%%%%%%%%%%%%%%%%%%%%

The normalisation of the short-distance cross sections is strongly
affected by the choice of the charm quark mass, the QCD coupling, the
renormalisation/factorisation scale $\mu$, and the parton distribution
functions. Varying the parameters in the range $1.35\,\mbox{GeV}\! <\!
m_c\! <\! 1.65\,\mbox{GeV}$, $m_c\! <\! \mu\!  <\! 4 m_c$, and
$150\,\mbox{MeV}\! <\! \Lambda^{(4)}\! < \!  250\,\mbox{MeV}$, the
normalisation of $d\hat{\sigma}(ij\to c\bar c[n])$ is altered by $\sim
\pm 50\%$ around the central value at $m_c = 1.5$~GeV, $\mu = 2 m_c$,
and $\Lambda^{(4)}=200$~MeV.\footnote{To study the $\alpha_s$
  dependence of the cross section, we use consistently adjusted sets
  of parton densities \cite{GRV,AV}.} Adopting e.g.\ the MRS(R2) set
of parton distributions \cite{MRS} and the corresponding value of
$\alpha_s$ decreases the short-distance cross sections by about a
factor of 2 as compared to the leading-order GRV parametrisation.
However, the values of the non-perturbative colour-octet matrix
elements as extracted from fits to the Tevatron data \cite{BEN97}
depend on the choice of $m_c$, $\alpha_s$, $\mu$ and the parton
distribution in approximately the opposite way such as to compensate
the change in the short-distance cross section. At leading-twist and
leading order in $\alpha_s$, the overall normalisation uncertainty of
the colour-octet contributions to $J/\psi$ photoproduction is thus in
the range of only about $\pm 10\%$, if the short-distance cross
sections are multiplied with non-perturbative matrix elements that
have been extracted from hadroproduction data using the same set of
input parameters. The long-distance factor of the colour-singlet cross
section $\langle {\cal O}_1^{J/\psi}(^3\!S_1) \rangle$ on the other
hand can be determined from the leptonic decay width and is not very
sensitive to the choice of parameters, up to unknown contributions
from next-to-next-to-leading-order QCD corrections.  Consequently, the
normalisation uncertainty of the short-distance cross section
$d\hat{\sigma}(ij\to c\bar c[^3\!S_1^{(1)}])$ is not compensated by a
change in the long-distance factor and the colour-singlet contribution
should be considered uncertain within a factor of two.
Next-to-leading order QCD corrections \cite{Kraemer} increase the
colour-singlet cross section by $\sim 20\% \!-\! 40\%$, depending in
detail on the choice of parameters, but do not affect the shape of the
$J/\psi$ energy distribution.

Given the large normalisation uncertainties in particular of the
colour-singlet contribution, no conclusive statement about the size of
the colour-octet matrix elements can be derived from the $J/\psi$
energy distribution in the region $z\;\simlt\;0.8$. On
the other hand, the dramatic increase of the colour-octet cross
section at larger $z$ is not supported by the data.  One should not
interpret this discrepancy as a failure of the NRQCD theory itself,
but rather as an artefact of our leading-order approximation in
$\alpha_s$ and $v^2$ for the colour-octet contributions. Close to the
boundary of phase space, for $z\;\simgt\; 0.75$, the shape of the
$z$-distribution cannot be predicted without resumming singular
higher-order terms in the velocity expansion \cite{BRW}. This
difficulty is exactly analogous to the well-known problem of
extracting the CKM matrix element $|V_{ub}|$ from the endpoint region
of the lepton energy distribution in semileptonic $B$ decay. To
constrain the colour-octet contributions from the $J/\psi$
$z$-distribution, the distribution would have to be averaged close to
the endpoint over a region {\em much larger} than $v^2\sim 0.25$.

The low-$z$ region is not expected to be sensitive to higher-order 
terms in the velocity expansion. Therefore, if the data could be 
extended to the low-$z$ region, an important resolved photon 
contribution should be visible, if the colour-octet matrix elements 
are not significantly smaller than assumed in Table~\ref{tab1}.

%%%%%%%%%%%%%%%%%%%%%%%%%%%%%%%%%%%%%%%%%%%%%%%%%%%%%%%%%%%%%%%%%%%%%%%%%%
\begin{figure}[p]
   \vspace{-2.3cm}
   \epsfysize=16.8cm
   \epsfxsize=12cm
   \centerline{\epsffile{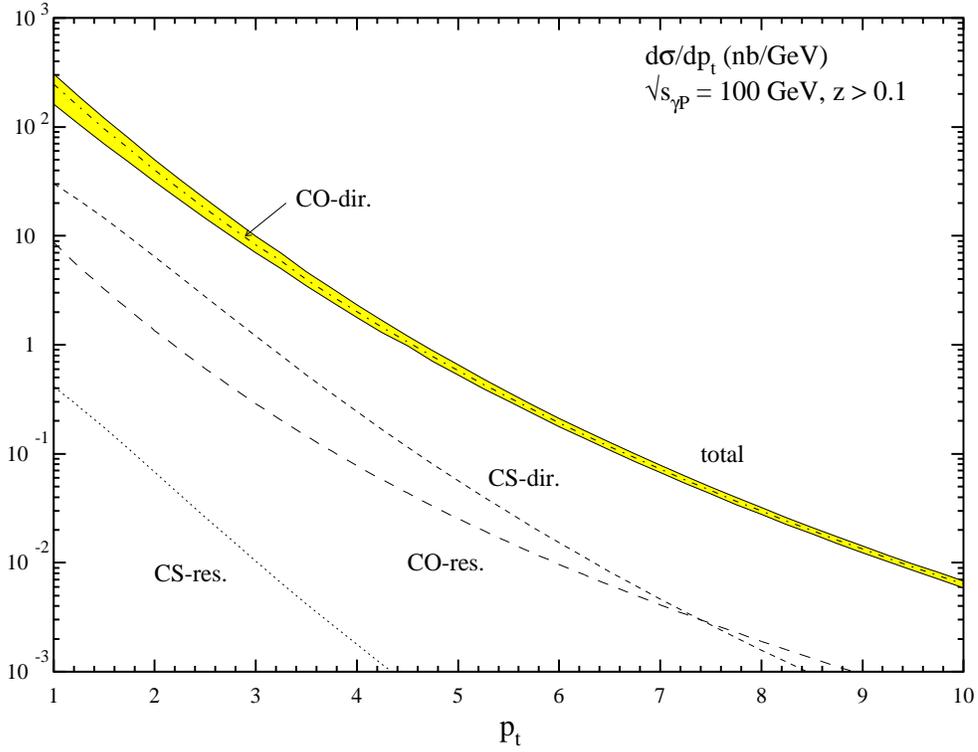}}
   \vspace*{-4cm}
\caption[dummy]{\small \label{sigmapt} Colour-singlet (CS) and colour-octet 
  (CO) contributions due to direct and resolved photons to the
  $J/\psi$ transverse momentum distribution $d\sigma/dp_t$ at the
  photon--proton centre-of-mass energy $\sqrt{s_{\gamma p}} =100\,$GeV;
  $z$ is integrated to its upper kinematic limit. Other specifications
  are as in Figure~\ref{sigmaz}.}
\end{figure}
%%%%%%%%%%%%%%%%%%%%%%%%%%%%%%%%%%%%%%%%%%%%%%%%%%%%%%%%%%%%%%%%%%%%%%%%%%

The $p_t$-distribution for inelastically produced $J/\psi$ is shown in
Figure~\ref{sigmapt} with a lower $z$-cut: $z>0.1$.  As discussed in
Section~\ref{theory} no upper cut in $z$ is necessary or advisable to
suppress the diffractive contribution, if the transverse momentum is
above about $1\,$GeV. With this definition the differential cross
section is dominated by colour-octet contributions, which exceed the
colour-singlet contribution by almost an order of magnitude, similar
to their significance in hadron--hadron collisions at fixed-target
energies \cite{FTpred3}.  Experimental data from HERA exist
only for $p_t<3\,$ GeV \cite{H1,ZEUSinelastic}. The data are presented
with a cut at $z<0.9$, in which case the differential cross section at
$p_t=1\,$GeV ($3\,$GeV) is found to be about a factor of 10 (2)
smaller than in Figure~\ref{sigmapt}.  The transverse momentum
distribution at $z < 0.9$ is adequately accounted for by the
colour-singlet channel, including next-to-leading-order corrections in
$\alpha_s$ \cite{Kraemer}. Diagrams with $t$-channel gluon exchange
lead to large $K$-factors that increase with increasing transverse
momentum and harden the $p_t$-spectrum of the colour-singlet channel
at NLO considerably.  We do not expect a similar strong impact of
next-to-leading order QCD corrections on the transverse momentum
distribution of the colour-octet cross sections. It would be
interesting to learn whether including all $z$ can lead to stringent
constraints on the size of the colour-octet matrix elements.  However,
in order to obtain an accurate theoretical prediction in the lower-$p_t$ 
region, $p_t\;\simlt\; 2$--$3\,$GeV, perturbative soft-gluon
resummation would have to be taken into account. We expect that 
soft-gluon resummation will be more important for the colour-octet 
processes, because there is no Sudakov form factor for radiation off the 
$c\bar{c}$ pair in the colour-singlet $^3\!S_1$ channel, for which there 
exists a colour dipole moment only.

\subsection{Decay angular distributions}

We now turn to the decay angular distributions, which constitute the
main result of this work. Below we present the $z$- and
$p_t$-dependence of the polar and azimuthal decay angular distribution
parameters $\lambda, \mu, \nu$ defined in (\ref{paramdef}), at a
typical HERA centre-of-mass energy of $\sqrt{s} = 100$~GeV.  The
quasi-real photons at HERA are actually not mono-energetic, but have a
distribution in energy given approximately by the
Weizs\"acker--Williams approximation. However, in general we have
found little energy dependence in the energy range relevant to HERA
(the only exception being the predictions in the recoil frame at
$z\;\simlt\; 0.3$) and thus considered a single energy.

Since the decay angular distribution parameters are normalised, the
dependence on parameters that affect the absolute normalisation of
cross sections, such as the charm quark mass, strong coupling, the
renormalisation/factorisation scale and parton distribution, cancels to
a large extent and does not constitute a significant uncertainty.

\begin{figure}[t]
   \vspace*{-1.5cm}
   \hspace*{0.25cm}
   \epsfysize=18cm
   \epsfxsize=14cm
   \centerline{\epsffile{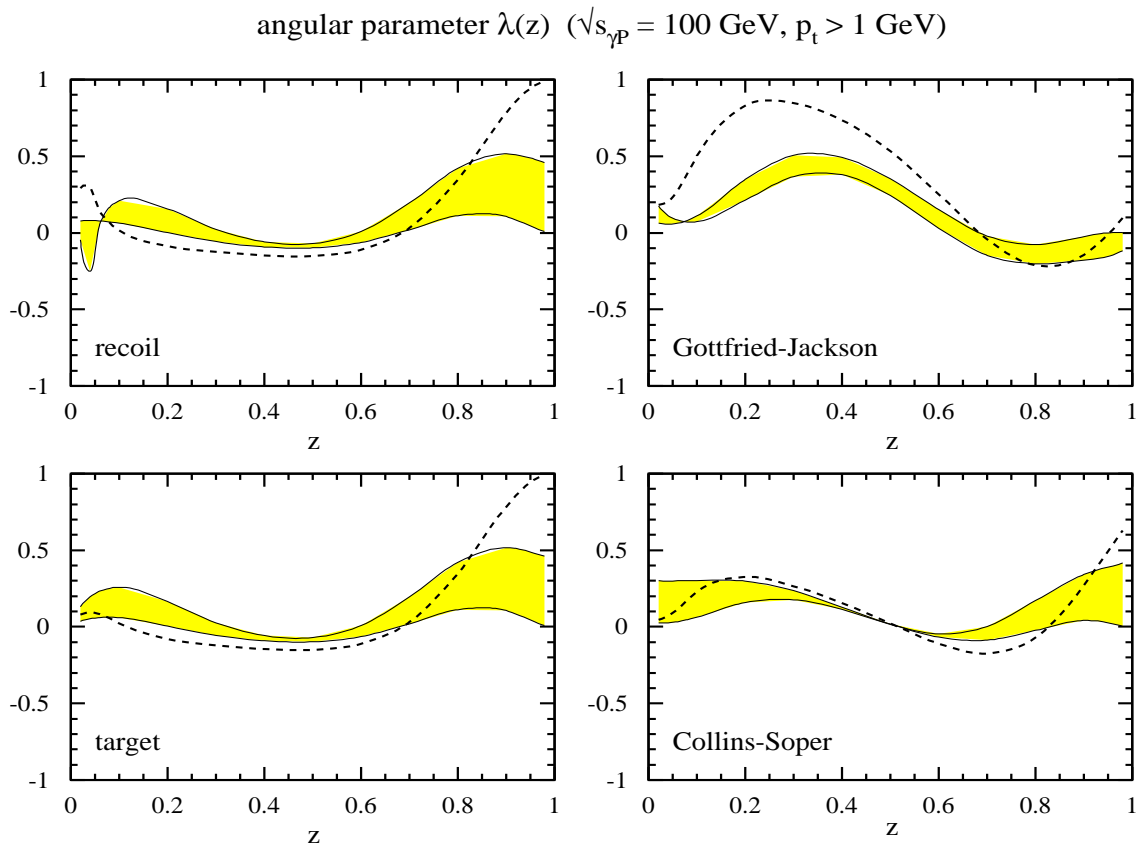}}
   \vspace*{-5cm}
\caption[dummy]{\small \label{anglez-l} Angular parameter $\lambda$ 
  of the decay angular distribution as a function of $z$.  Direct and
  resolved photons are included. The dashed line is the colour-singlet
  model prediction. The shaded area shows the NRQCD prediction bounded by
  the choice of parameters according to scenarios I and II. Other
  parameters are as in Figure~\ref{sigmaz}.}
\end{figure}

\begin{figure}[htb]
   \vspace*{-1.5cm}
   \hspace*{0.25cm}
   \epsfysize=18cm
   \epsfxsize=14cm
   \centerline{\epsffile{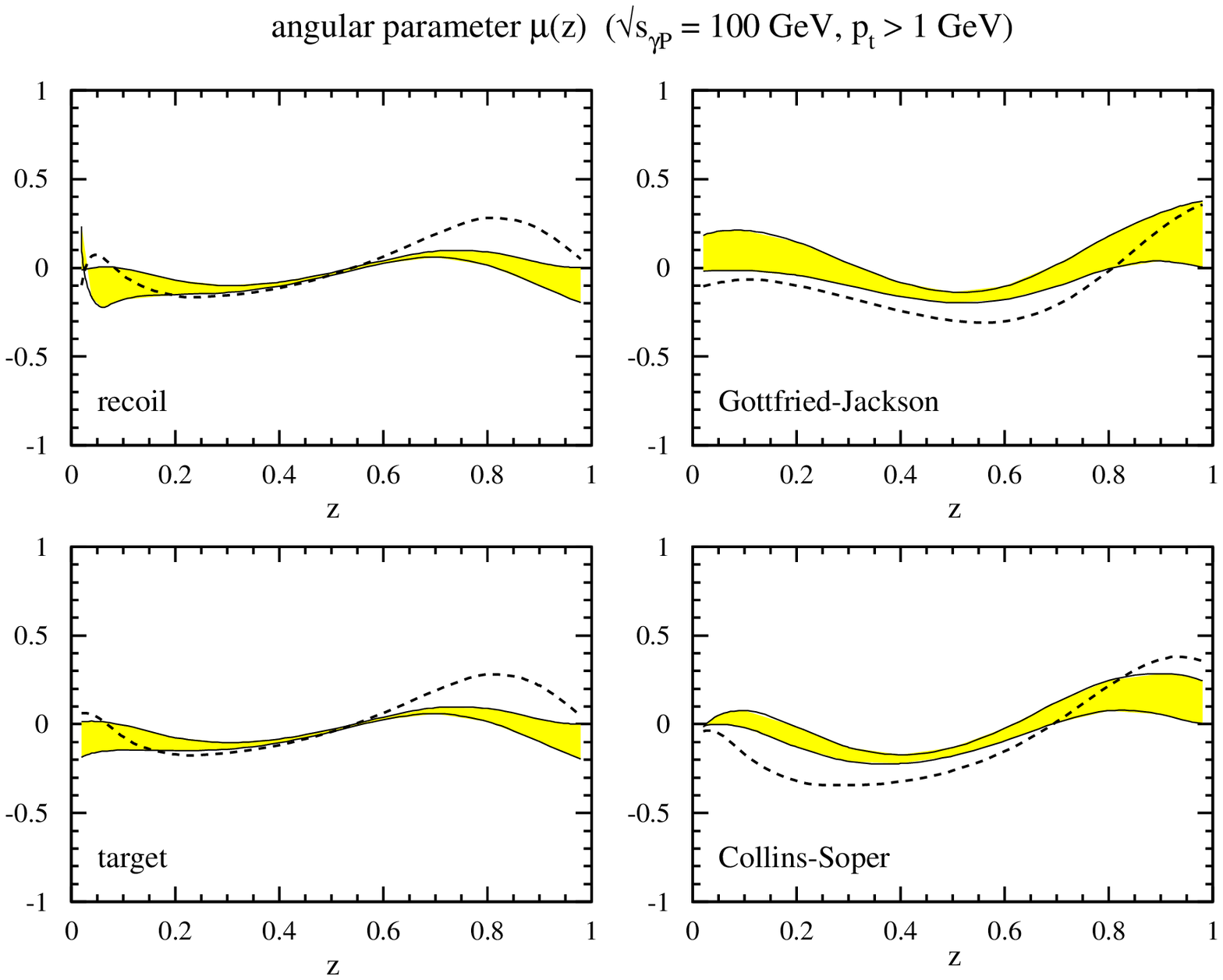}}
   \vspace*{-5cm}
\caption[dummy]{\small \label{anglez-m} Angular parameter $\mu$ 
  of the decay angular distribution as a function of $z$.  Direct and
  resolved photons are included. The dashed line is the colour-singlet
  model prediction. The shaded area shows the NRQCD prediction bounded by
  the choice of parameters according to scenarios I and II. Other
  parameters are as in Figure~\ref{sigmaz}.}
\end{figure}

\begin{figure}[htb]
   \vspace*{-1.5cm}
   \hspace*{0.25cm}
   \epsfysize=18cm
   \epsfxsize=14cm
   \centerline{\epsffile{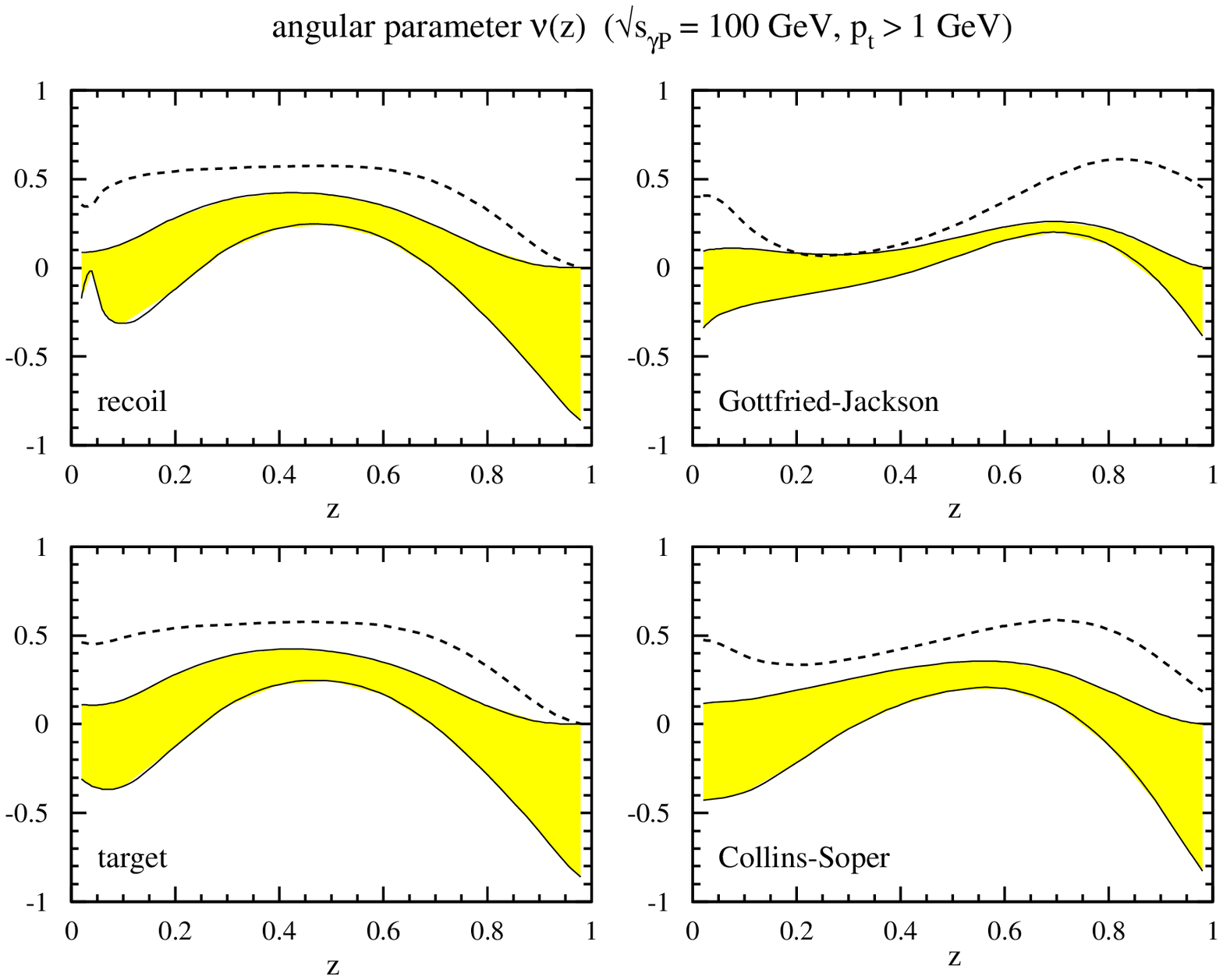}}
   \vspace*{-5cm}
\caption[dummy]{\small \label{anglez-n} Angular parameter $\nu$ 
  of the decay angular distribution as a function of $z$.  Direct and
  resolved photons are included. The dashed line is the colour-singlet
  model prediction. The shaded area shows the NRQCD prediction bounded by
  the choice of parameters according to scenarios I and II. Other
  parameters are as in Figure~\ref{sigmaz}.}
\end{figure}

The parameters $\lambda, \mu, \nu$ as function of $z$ are shown in
Figures~\ref{anglez-l}-\ref{anglez-n}, which include direct and
resolved photon contributions.  We computed the decay angular
distribution parameters in four commonly used frames (recoil or
$s$-channel helicity frame, Gottfried--Jackson frame, target frame and
Collins--Soper frame) defined in Appendix~\ref{pfs}. Each plot exhibits
the result from the colour-singlet channel alone (dashed line) and the
result after including the colour-octet contributions. The two solid
lines correspond to the two scenarios for the colour-octet matrix
elements discussed in Section~\ref{theory}. Recall that the $J/\psi$
is unpolarised, if it originates from a $c\bar{c}$ pair in a $^1\!S_0$
state. Thus, in scenario I the angular parameters $\lambda, \mu, \nu$
tend to zero in regions where the colour-octet processes dominate.
%%%%%%%%%%%%%%%%%%%%%%%%%%%%%%%%%%%%%%%%%%%%%%%%%%%%%%%%%%%%%%%%%%%%%%%%%%
%%%%%%%%%%%%%%%%%%%%%%%%%%%%%%%%%%%%%%%%%%%%%%%%%%%%%%%%%%%%%%%%%%%%%%%%%%
%%%%%%%%%%%%%%%%%%%%%%%%%%%%%%%%%%%%%%%%%%%%%%%%%%%%%%%%%%%%%%%%%%%%%%%%%%
%%%%%%%%%%%%%%%%%%%%%%%%%%%%%%%%%%%%%%%%%%%%%%%%%%%%%%%%%%%%%%%%%%%%%%%%%%

Inspecting Figure~\ref{anglez-l}, we note that in the recoil, target
and Collins--Soper frames $\lambda$ differs from the colour-singlet
prediction only in the endpoint region. The comparison looks different
in the Gottfried--Jackson frame: for $z\;\simlt\; 0.5$ the
colour-singlet channel yields large and positive values of $\lambda$,
while the colour-octet contributions yield almost unpolarised
$J/\psi$. The azimuthal parameter $\mu$ (Figure~\ref{anglez-m}) turns
out to be least interesting. We find that in all frames $\mu$ is
relatively flat and close to zero, for both the colour-singlet and
colour-octet contributions. The parameter $\nu$, on the other hand, is
very different in the colour-singlet channel and after inclusion of
colour-octet contributions, even in the intermediate region of $z$,
where the colour-singlet channel dominates. As can be seen from
Figure~\ref{anglez-n}, this difference is present in all frames and
seems to make $\nu$ the most useful parameter to find out about
the relative magnitude of colour-singlet and colour-octet
contributions experimentally. To determine $\nu$ one could measure
the decay angular distribution integrated over the polar angle (cf.
(\ref{paramdef})),
\begin{equation}
  \frac{d\sigma}{d\phi dy}
  \propto  
  1 + \frac{\lambda(y)}{3} + \frac{\nu(y)}{3} \cos 2\phi, 
\eeq
or project on $\nu$ as follows:
\begin{equation}
\nu(y) = \frac{8}{3} \cdot
         \frac{\int d\Omega \cos(2\phi) \frac{d\sigma}{d\Omega dy}}{
         \int d\Omega \left( 1 - \frac{5}{3}\cos^2\theta \right)
         \frac{d\sigma}{d\Omega dy}}.
\end{equation}
A distinctive signature of colour-octet contributions in the large-$z$ 
region could be of interest in connection with the difficulties 
in predicting the total cross section in the endpoint region. 
However, the endpoint region in Figures~\ref{anglez-l}-\ref{anglez-n} 
is not without problems either. The higher-order terms in the velocity 
expansion that need to be resummed close to the endpoint lead to 
a convolution of the $z$-distribution with certain non-perturbative 
shape functions. These shape functions depend on the production 
channel ($^3\!S_1^{(1)}$, $^1\!S_0^{(8)}$ and $^3\!P_J^{(8)}$) but 
they are the same for all density matrix elements in every given 
production channel. 
%%%%%%%%%%%%%%%%%%%%%%%%%%%%%%%%%%%%%%%%%%%%%%%%%%%%%%%%%%%%%%%%%%%%%%%%%%
\begin{figure}[t]
   \vspace*{-1.5cm}
   \hspace*{0.25cm}
   \epsfysize=18cm
   \epsfxsize=14cm
   \centerline{\epsffile{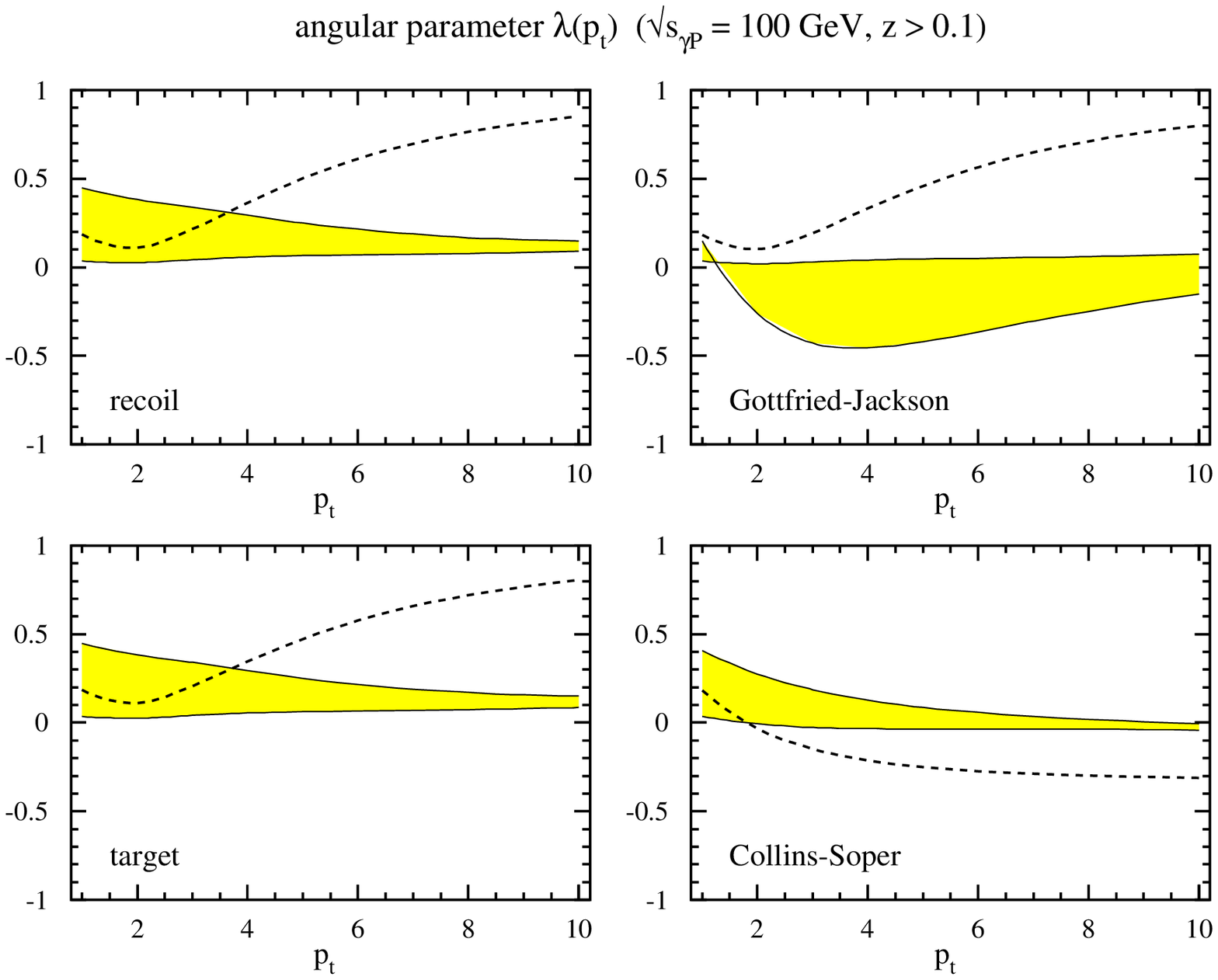}}
   \vspace*{-5cm}
\caption[dummy]{\small \label{anglept-l} Angular parameter $\lambda$ 
  of the decay angular distribution as a function of $p_t$.  Direct and
  resolved photons are included. The dashed line is the colour-singlet
  model prediction. The shaded area shows the NRQCD prediction bounded by
  the choice of parameters according to scenarios I and II. Other
  parameters are as in Figure~\ref{sigmaz}.}
\end{figure}
%%%%%%%%%%%%%%%%%%%%%%%%%%%%%%%%%%%%%%%%%%%%%%%%%%%%%%%%%%%%%%%%%%%%%%%%%%
\begin{figure}[htb]
   \vspace*{-1.5cm}
   \hspace*{0.25cm}
   \epsfysize=18cm
   \epsfxsize=14cm
   \centerline{\epsffile{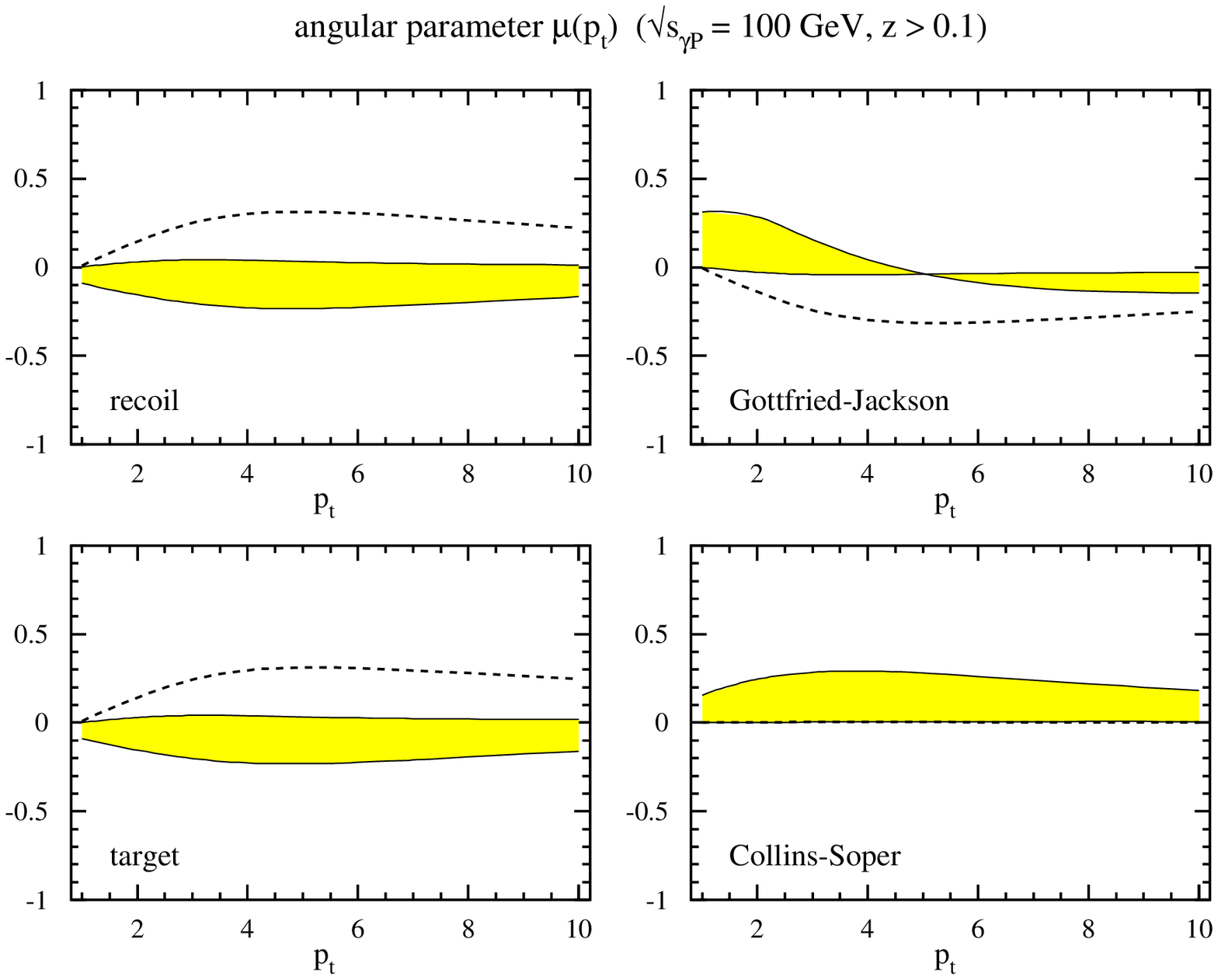}}
   \vspace*{-5cm}
\caption[dummy]{\small \label{anglept-m} Angular parameter $\mu$ 
  of the decay angular distribution as a function of $p_t$.  Direct and
  resolved photons are included. The dashed line is the colour-singlet
  model prediction. The shaded area shows the NRQCD prediction bounded by
  the choice of parameters according to scenarios I and II. Other
  parameters are as in Figure~\ref{sigmaz}.}
\end{figure}
%%%%%%%%%%%%%%%%%%%%%%%%%%%%%%%%%%%%%%%%%%%%%%%%%%%%%%%%%%%%%%%%%%%%%%%%%%
\begin{figure}[htb]
   \vspace*{-1.5cm}
   \hspace*{0.25cm}
   \epsfysize=18cm
   \epsfxsize=14cm
   \centerline{\epsffile{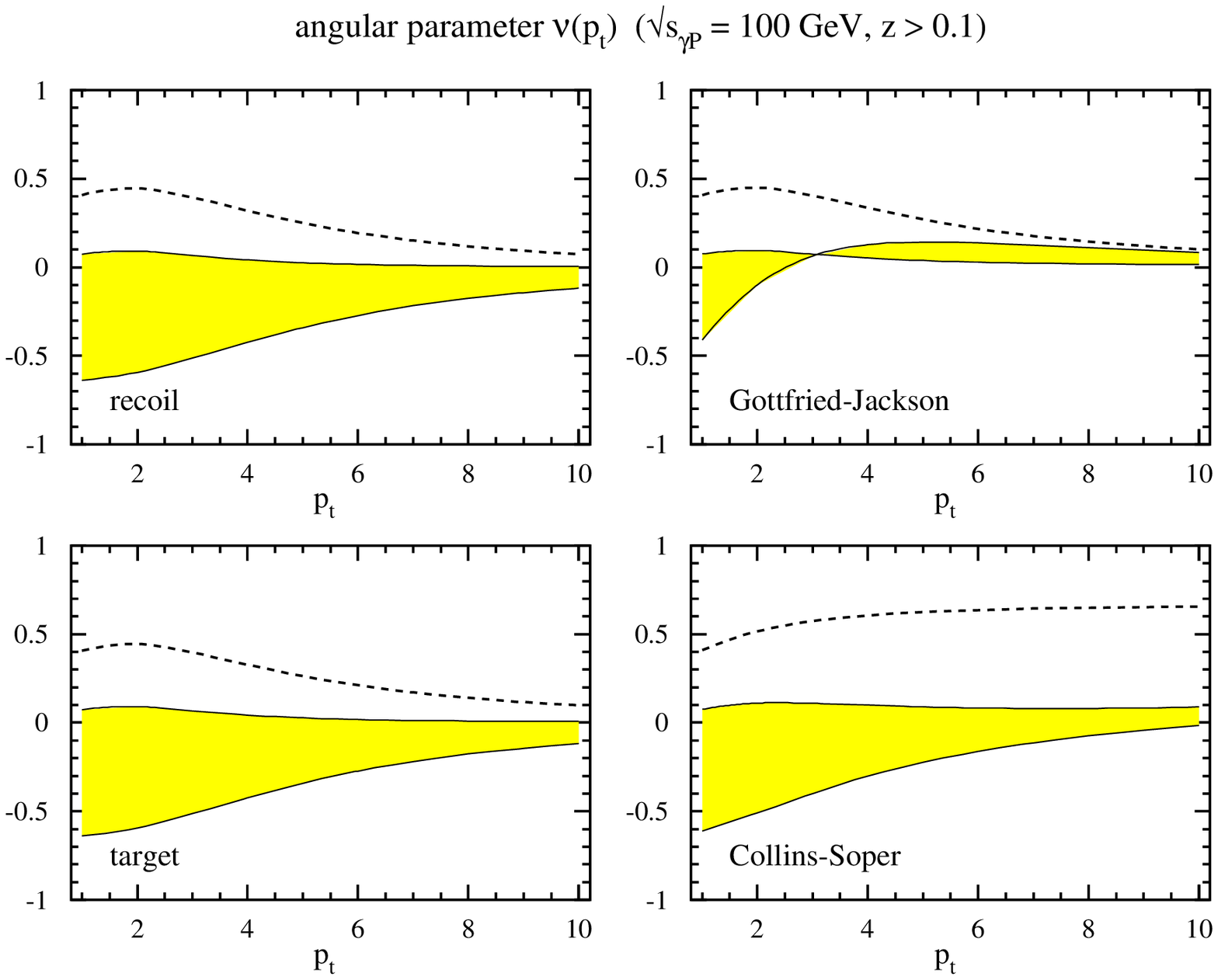}}
   \vspace*{-5cm}
\caption[dummy]{\small \label{anglept-n} Angular parameter $\nu$ 
  of the decay angular distribution as a function of $p_t$.  Direct and
  resolved photons are included. The dashed line is the colour-singlet
  model prediction. The shaded area shows the NRQCD prediction bounded by
  the choice of parameters according to scenarios I and II. Other
  parameters are as in Figure~\ref{sigmaz}.}
\end{figure}
%%%%%%%%%%%%%%%%%%%%%%%%%%%%%%%%%%%%%%%%%%%%%%%%%%%%%%%%%%%%%%%%%%%%%%%%%%
As a consequence, while the energy distribution itself 
depends on these shape functions, the {\em moments} in 
$z$ of the normalised angular parameters depend only on the difference of 
the shape functions in the various production channels. Since we do 
expect such differences, especially 
between the colour-singlet and the colour-octet channels (due to 
the different properties with respect to soft gluon radiation), 
and since we are interested in the $z$-distribution rather than its moments, 
the predictions for the angular parameters in the endpoint 
region still depend on these shape functions. However, 
this dependence is strong only if the angular parameter varies 
strongly in the endpoint region and disappears entirely if its 
distribution is flat.

In Figures~\ref{anglept-l}-\ref{anglept-n} we present the same 
analysis for the $p_t$-distribution. 
We note that $z$ is integrated up to its kinematic maximum. As a 
consequence the cross section is colour-octet dominated and the 
colour-singlet contribution plays no role for the solid curves. 
Since the colour-octet cross section is dominated by the large-$z$ 
region, the solid curves are entirely determined by the polarisation 
yield of octet mechanisms at large $z$. Contrary to the situation 
of hadroproduction at the Tevatron, where one expects large transverse 
polarisation due to gluon fragmentation into colour-octet $c\bar{c}$ 
pairs \cite{CDFpred}, the photoproduction cross section tends to be 
unpolarised in the $p_t$ region considered here. Therefore, the 
$p_t$-distributions do not discriminate between the theoretical prediction 
based on NRQCD and that of the colour evaporation model, which 
always predicts unpolarised $J/\psi$. 

On the other hand, the transverse momentum distribution seems to prove 
very useful to determine the relative magnitude of colour-singlet 
and colour-octet contributions: If the cross section is dominated 
by the colour-singlet channel, large and positive values of the polar 
angular parameter $\lambda$ (Figure~\ref{anglept-l}) are expected for 
$p_t\;\simgt\;5$~GeV in the recoil, Gottfried--Jackson and target frames. 
A similar distinctive difference between colour-singlet and colour-octet 
channels is visible in the azimuthal parameter $\nu$ as defined in the 
Collins--Soper frame, see Figure~\ref{anglept-n}. 

The unique transverse polarisation signature of gluon fragmentation 
could possibly be made visible in the resolved photon contribution. 
If the direct photon contribution is reduced by a cut on the high-$z$ 
region, the resolved cross section is dominated by 
$g\to c\bar{c}[^3\!S_1^{(8)}]$ at large transverse 
momentum \cite{Kniehl}. As for 
hadroproduction \cite{ChoL,BEN97}, 
we expect that $p_t\;\simgt\; (5$--$10)\,$GeV is necessary to suppress 
sufficiently the other colour-octet channels. 

\section{Conclusion \label{summary}}
%%%%%%%%%%%%%%%%%%%%%%%%%%%%%%%%%%%%%%%%%%%%%%%%%%%%%%%%%%%%%%%%%%%%%%

We presented an analysis of the full polar and azimuthal decay 
angular distributions of inelastically photo-produced $J/\psi$ 
based on the NRQCD factorisation approach to quarkonium production. 
The primary motivation of this study is to find observables 
in addition to angular-integrated differential cross sections, 
which are sensitive to different theories and models of quarkonium 
production. A particular emphasis is on clarifying the role 
of colour-octet contributions suggested by NRQCD and other 
quarkonium production processes in comparison with the colour-singlet 
model, which can be considered as a successful description of 
photoproduction as far as present, limited, data on energy 
and transverse momentum distributions is concerned.

Assuming NRQCD long-distance parameters 
as constrained by other $J/\psi$ production processes such as 
in hadroproduction and $B$ decay, we have found that the 
azimuthal decay angle distribution as a function of $z$ or 
$p_t$ would be extremely useful for discriminating between  
the colour-singlet model and NRQCD and, to a lesser extent, 
the colour evaporation model. We also noted that 
transverse-momentum distributions integrated over all 
energy fraction $z$ are colour-octet dominated and  
could give meaningful constraints on colour-octet matrix 
elements from the angular integrated rate as well as the 
decay angle dependence.

While this paper was being written, Fleming and Mehen \cite{FLE97}
presented a study of leptoproduction of $J/\psi$ in NRQCD, 
which is complementary to our photoproduction analysis. 
Contrary to photoproduction, the leading-twist partonic diagrams 
at $O(\alpha\alpha_s)$ can be sensibly compared with 
a total leptoproduction cross section for large enough 
photon virtuality $Q^2$. 
Fleming and Mehen computed the polar angle distribution 
due to these leading-order mechanisms and also find interesting 
tests of NRQCD.\\

\noindent {\bf Acknowledgements.} We would like to thank J.G.~K\"orner 
for useful discussions. M.B. wishes to thank John Collins 
for supplying a FORTRAN integration routine.

\appendix
%%%%%%%%%%%%%%%%%%%%%%%%%%%%%%%%%%%%%%%%%%%%%%%%%%%%%%%%%%%%%%%%%%%%%%
\section{Polarisation frames}
\label{pfs}

We collect here the covariant expressions for polarisation vectors 
in the four commonly used frames, following \cite{LAM78}. (Note 
that the metric $g_{\mu\nu}=\mbox{diag}(-1,1,1,1)$ is used there.) 

Let $p_\gamma$ be the photon momentum, $p_p$ the proton momentum, 
$P$ the momentum of quarkonium $\psi$ and 
$s=(p_\gamma+p_p)^2$. We define the auxiliary vectors 
\begin{eqnarray}
A=p_\gamma+p_p \qquad \tilde{A}^\mu=A^\mu-\frac{A\cdot P P^\mu}{M^2},
\\
B=p_\gamma-p_p \qquad \tilde{B}^\mu=B^\mu-\frac{B\cdot P P^\mu}{M^2},
\end{eqnarray}
where $M$ is the $\psi$ mass. (We take $M=2 m_c$ in the 
analysis. The proton mass will be neglected in the 
following.) Note that $\tilde{A}$, $\tilde{B}$ are three-vectors in 
the $\psi$-rest frame. 
We then define a coordinate system as follows:
\begin{enumerate}
\item Choose $Z^\mu=\alpha_z\tilde{A}^\mu+\beta_z\tilde{B}^\mu$, 
with $Z^2=-1$.
\item Define $X^\mu=\alpha_x\tilde{A}^\mu+\beta_x\tilde{B}^\mu$ in the 
plane spanned by $\tilde{A}$, $\tilde{B}$, orthogonal to $Z$ and 
normalised: $X\cdot Z=0$, $X^2=-1$.
\item Take $Y$ to complete a right-handed coordinate system in the 
$\psi$-rest frame,
\begin{equation}
Y^\mu=\frac{1}{M}\epsilon^{\mu\alpha\beta\gamma} 
P_\alpha X_\beta Z_\gamma 
\end{equation}
($\epsilon^{0123}=1$). The sign ambiguity in $\alpha_x$, $\beta_x$ 
left in the second step is fixed by requiring $\vec{Y}$ to point in 
the direction of $\vec{p_\gamma}\times (-\vec{p}_p)$ in the 
$\psi$-rest frame, which requires $\alpha_z\beta_x-\alpha_x\beta_z>0$.
\end{enumerate}
The four commonly used polarisation frames are then specified by 
the choice of $Z$. In the {\em recoil} (or $s$-channel helicity) frame, 
the $Z$-axis is defined as the direction of the $\psi$ three-momentum 
in the hadronic centre-of-mass frame, that is $\vec{Z}=-\vec{A}/|\vec{A}|$ 
in the $\psi$-rest frame. In the {\em Gottfried--Jackson} frame 
$\vec{Z}=\vec{p}_\gamma/|\vec{p}_\gamma|$ and in the {\em target} frame 
$\vec{Z}=-\vec{p}_p/|\vec{p}_p|$. In the {\em Collins--Soper} frame 
the $Z$-axis bisects the angle between $\vec{p}_\gamma$ and $(-\vec{p}_p)$, 
i.e. $\vec{Z}\propto \vec{p}_\gamma/|\vec{p}_\gamma|+(-\vec{p}_p)/
|\vec{p}_p|$. (All three-vectors refer to the $\psi$-rest frame.) 
We then find the covariant expressions for the coordinate axes from 
the following expressions for $\alpha_{z,x}$, $\beta_{z,x}$:\\

Recoil frame:
\begin{equation}
\alpha_z=-\frac{M}{\sqrt{(A\cdot P)^2-M^2 s}} \qquad 
\beta_z=0
\end{equation}
\begin{eqnarray}
\alpha_x &=& \frac{A\cdot P B\cdot P}{\sqrt{s \,((A\cdot P)^2-M^2 s) 
((A\cdot P)^2-(B\cdot P)^2-M^2 s)}}
\\
\beta_x &=& -\frac{\sqrt{(A\cdot P)^2-M^2 s}}{\sqrt{s\,  
((A\cdot P)^2-(B\cdot P)^2-M^2 s)}}
\end{eqnarray}

Gottfried--Jackson frame:
\begin{equation}
\alpha_z=\beta_z=\frac{M}{A\cdot P+B\cdot P} 
\end{equation}
\begin{eqnarray}
\alpha_x &=& -\frac{(B\cdot P)^2+A\cdot P B\cdot P+M^2 s}
{(A\cdot P+B\cdot P)\sqrt{s \,((A\cdot P)^2-(B\cdot P)^2-M^2 s)}}
\\
\beta_x &=& \frac{(A\cdot P)^2+A\cdot P B\cdot P-M^2 s}
{(A\cdot P+B\cdot P)\sqrt{s \,((A\cdot P)^2-(B\cdot P)^2-M^2 s)}}
\end{eqnarray}

Target frame:
\begin{equation}
\alpha_z=-\beta_z=-\frac{M}{A\cdot P-B\cdot P} 
\end{equation}
\begin{eqnarray}
\alpha_x &=& -\frac{(B\cdot P)^2-A\cdot P B\cdot P+M^2 s}
{(A\cdot P-B\cdot P)\sqrt{s \,((A\cdot P)^2-(B\cdot P)^2-M^2 s)}}
\\
\beta_x &=& -\frac{(A\cdot P)^2-A\cdot P B\cdot P-M^2 s}
{(A\cdot P-B\cdot P)\sqrt{s \,((A\cdot P)^2-(B\cdot P)^2-M^2 s)}}
\end{eqnarray}

Collins--Soper frame:
\begin{equation}
\alpha_z=-\frac{B\cdot P}{\sqrt{s\,((A\cdot P)^2-(B\cdot P)^2)}} \qquad 
\beta_z=\frac{A\cdot P}{\sqrt{s \,((A\cdot P)^2-(B\cdot P)^2)}}
\end{equation}
\begin{eqnarray}
\alpha_x &=& -\frac{M\,A\cdot P}
{\sqrt{((A\cdot P)^2-(B\cdot P)^2) \,((A\cdot P)^2-(B\cdot P)^2-M^2 s)}}
\\
\beta_x &=& \frac{M\,B\cdot P}
{\sqrt{((A\cdot P)^2-(B\cdot P)^2) \,((A\cdot P)^2-(B\cdot P)^2-M^2 s)}}
\end{eqnarray}
\vspace*{0.2cm}

\noindent We note 
that $A\cdot P+B\cdot P=(P_t^2+M^2)/z$ and 
$A\cdot P-B\cdot P=s z$. Finally, the polarisation vectors are given by
\begin{equation}
\epsilon^\mu(\lambda=0)=Z^\mu \qquad 
\epsilon^\mu(\lambda=\pm 1)=\frac{1}{\sqrt{2}}(\mp X^\mu-i Y^\mu).
\end{equation}

\section{Density matrices}
\label{dmes}

The density matrices for all processes considered in the paper 
are given in this Appendix. The results given for the resolved 
photon process apply equally to $J/\psi$ production in hadron-hadron 
collisions and have been used in \cite{BEN97}. 
The functions $F,a,b,c,d$ below are related to $A,B,C,D$ of 
(\ref{rhogeneral}) as $A = Fa$ etc., and we suppressed all 
sub- and superscripts. For the partonic process 
$1+2\to 3+4$, the Mandelstam invariants are $\hat{s}=(p_1+p_2)^2$, 
$\hat{t}=(p_3-p_1)^2$, $\hat{u}=(p_3-p_2)^2$.

\subsection{Direct-photon processes}

\noindent $\gamma + g
\rightarrow c\bar c \left[ ^3\!S_1^{(1)} \right] + g$:\\[-1.2cm]

\beqa
  F &=& \frac{16 M (4\pi)^3 \alpha\alpha_s^2 e_c^2
          \langle {\cal O}_1^{J/\psi}(^3\!S_1) \rangle}{27
          [ (\sh-M^2) (\th-M^2) (\uh-M^2) ]^2} \\
  a &=& - (\sh^2 + \sh\th + \th^2)^2 + M^2 (2\sh^2 + \sh\th + 2\th^2)(\sh+\th)
          - M^4 (\sh^2 + \sh\th + \th^2)
          \label{a-singlet} \\
  b &=& -2(\sh^2 + \th^2) \\
  c &=& -2(\sh^2 + \uh^2) \\
  d &=& -2\sh^2 \label{d-singlet}
\eeqa

\noindent $\gamma + g
\rightarrow c\bar c \left[ ^3\!S_1^{(8)} \right] + g$: 
$a,b,c,d$ are the same as (\ref{a-singlet})--(\ref{d-singlet}),
$F$ is multiplied by
\beq
  \frac{15}{8} \cdot
  \frac{\langle {\cal O}_8^{J/\psi}(^3\!S_1) \rangle}{
  \langle {\cal O}_1^{J/\psi}(^3\!S_1) \rangle}
\eeq

\noindent $\gamma + g
\rightarrow c\bar c \left[ ^1\!S_0^{(8)} \right] + g$:\\[-1.2cm]

\beqa
  Fa &=& -\frac{2 (4\pi)^3 \alpha\alpha_s^2 e_c^2 
           \langle {\cal O}_8^{J/\psi}(^1\!S_0) \rangle}{
           M \th [ (\sh-M^2) (\th-M^2) (\uh-M^2) ]^2}
           \nonumber \\
     & & \times \sh\uh
           \left\{ [\uh^2-M^2(\uh-M^2)]^2 - 2\sh\th (\uh-M^2)^2 + 
           \sh^2 \th^2 \right\} \\
  b  &=& c = d = 0 
\eeqa

\noindent $\gamma + g
\rightarrow c\bar c \left[ ^3\!P_J^{(8)} \right] + g$:\\[-1.2cm]

\beqa
  F &=& -\frac{24 (4\pi)^3 \alpha\alpha_s^2 e_c^2 
          \langle {\cal O}_8^{J/\psi}(^3\!P_0) \rangle}{
          \th^2 [ M (\sh-M^2) (\th-M^2) (\uh-M^2) ]^3} \\
  a &=& \sh^2\th^2 (\sh+\th)^2 (\sh^2+\sh\th+\th^2)^2 
          \nonumber \\[-0.2cm]
    & & \mbox{} - M^2
          \sh\th(\sh^7 + 8\sh^6\th + 16\sh^5\th^2 + 16\sh^4\th^3 
          + 8\sh^3\th^4 - 3\sh\th^6 - 2\th^7)
          \nonumber \\[-0.2cm]
    & & \mbox{} + M^4
          \th(4\sh^7 + 28\sh^6\th + 48\sh^5\th^2 + 41\sh^4\th^3 
          + 18\sh^3\th^4 - 4\sh\th^6 + \th^7)
          \nonumber \\[-0.2cm]
    & & \mbox{} - M^6
          \th(12\sh^6 + 64\sh^5\th + 99\sh^4\th^2 + 80\sh^3\th^3
          + 33\sh^2\th^4 + 3\sh\th^5 + 3\th^6)
          \nonumber \\[-0.2cm]
    & & \mbox{} + M^8
          \th(22\sh^5 + 88\sh^4\th + 114\sh^3\th^2 + 67\sh^2\th^3 
          + 14\sh\th^4 + 3\th^5)
          \nonumber \\[-0.2cm]
    & & \mbox{} - M^{10}
          \th(22\sh^4 + 68\sh^3\th + 61\sh^2\th^2 + 16\sh\th^3 + \th^4)
          \nonumber \\[-0.2cm]
    & & \mbox{} + 2M^{12} \sh\th(6\sh^2 + 13\sh\th + 5\th^2)
          - 3M^{14} \sh\th(\sh+\th)
          \\
  b &=& -2(\sh+\th)^4 (2\sh^4+\sh^2\th^2+2\th^4)
          \nonumber \\[-0.2cm]
    & & \mbox{} + 2M^2 
          (6\sh^7 + 19\sh^6\th + 21\sh^5\th^2 + 17\sh^4\th^3 + 21\sh^3\th^4
          + 28\sh^2\th^5 + 22\sh\th^6 + 6\th^7)
          \nonumber \\[-0.2cm]
    & & \mbox{} - 2M^4
          (6\sh^6 + 9\sh^5\th - 15\sh^4\th^2 - 24\sh^3\th^3 - 4\sh^2\th^4 
          + 15\sh\th^5 + 5\th^6)
          \nonumber \\[-0.2cm]
    & & \mbox{} + 2M^6 
          (2\sh^5 - 10\sh^4\th - 45\sh^3\th^2 - 37\sh^2\th^3 - 3\sh\th^4
          + \th^5)
          \nonumber \\[-0.2cm]
    & & \mbox{} + 8M^8 \sh\th(3\sh^2 + 7\sh\th + 2\th^2)
          - 8M^{10} \sh\th(\sh + \th)
          \\
  c &=& 2\th^2(2\sh^6 + 6\sh^5\th + 5\sh^4\th^2 - 7\sh^2\th^4 - 6\sh\th^5 
          - 2\th^6)
          \nonumber \\[-0.2cm]
    & & \mbox{} + 2M^2 
          \th^2( - 6\sh^5 - 9\sh^4\th + 2\sh^3\th^2 + 21\sh^2\th^3 
          + 22\sh\th^4 + 8\th^5)
          \nonumber \\[-0.2cm]
    & & \mbox{} - 2M^4
          (2\sh^6 + 6\sh^5\th - 11\sh^4\th^2 - 20\sh^3\th^3 + 7\sh^2\th^4 
          + 27\sh\th^5 + 11\th^6)
          \nonumber \\[-0.2cm]
    & & \mbox{} + 2M^6 
          (6\sh^5 + 15\sh^4\th - 12\sh^3\th^2 - 19\sh^2\th^3 
          + 8\sh\th^4 + 4\th^5)
          \nonumber \\[-0.2cm]
    & & \mbox{} - 4M^8 (3\sh^4 + 6\sh^3\th - 4\sh^2\th^2 - 3\sh\th^3 - 2\th^4)
          \nonumber \\[-0.2cm]
    & & \mbox{} + 4M^{10} (\sh^3 + 2\sh^2\th - \sh\th^2 - 2\th^3) 
          - 2M^{12} \th(\sh - \th)
          \\
  d &=& 2\th^2(\sh^6 + 2\sh^5\th - 6\sh^3\th^3 - 10\sh^2\th^4 
          - 7\sh\th^5 - 2\th^6)
          \nonumber \\[-0.2cm]
    & & \mbox{} - M^2
          (4\sh^7 + 14\sh^6\th + 30\sh^5\th^2 + 29\sh^4\th^3 + 4\sh^3\th^4
          - 33\sh^2\th^5 - 38\sh\th^6 - 14\th^7) 
          \nonumber \\[-0.2cm]
    & & \mbox{} + 2M^4 
          (6\sh^6 + 21\sh^5\th + 43\sh^4\th^2 + 43\sh^3\th^3 
          + 13\sh^2\th^4 - 15\sh\th^5 - 9\th^6)
          \nonumber \\[-0.2cm]
    & & \mbox{} - M^6 (12\sh^5 + 45\sh^4\th + 92\sh^3\th^2 + 78\sh^2\th^3 
          + 6\sh\th^4 - 9\th^5)
          \nonumber \\[-0.2cm]
    & & \mbox{} + 2M^8 \sh(2\sh^3 + 11\sh^2\th + 22\sh\th^2 + 9\th^3)
          - M^{10} \th(5\sh^2 + 6\sh\th + \th^2)
\eeqa

\noindent $\gamma + q
\rightarrow c\bar c \left[ ^3\!S_1^{(8)} \right] + q$:\\[-1.2cm] 

\beqa
  F &=& \frac{(4\pi)^3 \alpha\alpha_s^2 e_q^2
          \langle {\cal O}_8^{J/\psi}(^3\!S_1) \rangle}
          {9 M^3 \sh \uh} \\
  a &=& \sh^2+\uh^2+2 M^2\th \\
  b &=& 4 \\
  c &=& 8 \\
  d &=& 4, 
\eeqa
where $e_q$ is the electric charge of the light quark $q$.\\ 

\noindent $\gamma + q
\rightarrow c\bar c \left[ ^1\!S_0^{(8)} \right] + q$:\\[-1.2cm]

\beqa
  Fa &=& \frac{4 (4\pi)^3 \alpha\alpha_s^2 e_c^2 
           \langle {\cal O}_8^{J/\psi}(^1\!S_0) \rangle}{
           9 M \th (\th-M^2)^2}\,\left\{\sh^2+\uh^2\right\} \\
  b  &=& c = d = 0 
\eeqa

\noindent $\gamma + q
\rightarrow c\bar c \left[ ^3\!P_J^{(8)} \right] + q$:\\[-1.2cm]

\beqa
  F &=& \frac{16 (4\pi)^3 \alpha\alpha_s^2 e_c^2 
          \langle {\cal O}_8^{J/\psi}(^3\!P_0) \rangle}{
          3 M^3 \th^2 (\th-M^2)^3} \\
  a &=& \th (\th-M^2) (\sh^2+\uh^2+2 M^2\th+2 M^4) \\
  b &=& -8 (\sh^2+\sh\th+M^2\th) \\
  c &=& 8 (\th^2-M^4) \\
  d &=& 4 (\th^2-2 M^2\sh-M^2\th)
\eeqa

\subsection{Resolved-photon processes}

\noindent $g + g
\rightarrow c\bar c \left[ ^3S_1^{(1)} \right] + g$: $a,b,c,d$ are 
the same as (\ref{a-singlet})--(\ref{d-singlet}),
$F$ is multiplied by
\beq
  \frac{5}{96} \cdot \frac{\alpha_s}{\alpha e_c^2}
\eeq

\noindent $g + g
\rightarrow c\bar c \left[ ^3S_1^{(8)} \right] + g$:\\[-1.2cm]

\beqa
  F &=&  \frac{(4\pi\alpha_s)^3 \langle {\cal O}_8^{J/\psi}(^3\!S_1) 
           \rangle}{144 M^3
          [ (\sh-M^2) (\th-M^2) (\uh-M^2) ]^2} \,\left\{
          27(\sh\th+\sh\uh+\th\uh)-19 M^4
          \right\}\\
  a &=& (\sh^2+\sh\th+\th^2)^2 - M^2 (2 \sh^2+\sh\th+2\th^2) (\sh+\th)
          + M^4 (\sh^2+\sh\th+\th^2) \\
  b &=& 2 (\sh^2+\th^2) \\
  c &=& 2 (\sh^2+\uh^2)\\
  d &=& 2 \sh^2
\eeqa

\noindent $g + g
\rightarrow c\bar c \left[ ^1S_0^{(8)} \right] + g$:\\[-1.2cm]

\beqa
  Fa &=& -\frac{5 (4\pi\alpha_s)^3
           \langle {\cal O}_8^{J/\psi}(^1S_0) \rangle}{
           48M \sh\th\uh [ (\sh-M^2)(\th-M^2)(\uh-M^2) ]^2}
           \nonumber \\
     & & \times \left\{\sh^2 (\sh-M^2)^2 + \sh\th\uh (M^2-2\sh) 
          + \th^2 \uh^2 \right\}
           \nonumber \\
     & & \times \left\{ (\sh^2-M^2 \sh+M^4)^2 - \th\uh (2\th^2
          +3\th\uh+2\uh^2) \right\} \\
  b  &=& c = d = 0 
\eeqa

\noindent $g + g
\rightarrow c\bar c \left[ ^3P_J^{(8)} \right] + g$:\\[-1.2cm]

\beqa
  F &=&  \frac{5 (4\pi\alpha_s)^3 
          \langle {\cal O}_8^{J/\psi}(^3\!P_0) \rangle}{4 
          \sh^2\th^2\uh^2 [ M (\sh-M^2) (\th-M^2) (\uh-M^2) ]^3} \\
  a &=& \sh\th\uh \,\Big\{\sh^{10}\th + 5\sh^9\th^2 + 14\sh^8\th^3 
          + 26\sh^7\th^4 + 35\sh^6\th^5 + 35\sh^5\th^6 
          + 26\sh^4\th^7 + 14\sh^3\th^8 
          \nonumber \\[-0.2cm]
    & & \mbox{} + 5\sh^2\th^9 + \sh\th^{10} 
          \nonumber \\[-0.2cm] 
    & & \mbox{} - M^2(\sh^{10} + 10\sh^9\th + 26\sh^8\th^2 
          + 40\sh^7\th^3 + 45\sh^6\th^4 + 44\sh^5\th^5 + 45\sh^4\th^6 
          \nonumber \\[-0.2cm]
    & & \mbox{} + 40\sh^3\th^7 + 26\sh^2\th^8 + 10\sh\th^9 + \th^{10})
          \nonumber \\[-0.2cm] 
    & & \mbox{} + M^4(5\sh^9 + 40\sh^8\th + 84\sh^7\th^2 + 106\sh^6\th^3 + 
          103\sh^5\th^4 + 103\sh^4\th^5 + 106\sh^3\th^6 
          \nonumber \\[-0.2cm]
    & & \mbox{} + 84\sh^2\th^7 + 40\sh\th^8 + 5\th^9)
          \nonumber \\[-0.2cm] 
    & & \mbox{} - M^6(16\sh^8 + 104\sh^7\th + 215\sh^6\th^2 + 
          291\sh^5\th^3 + 316\sh^4\th^4 + 291\sh^3\th^5 + 
          215\sh^2\th^6          
          \nonumber \\[-0.2cm]
    & & \mbox{} + 104\sh\th^7 + 16\th^8)
          \nonumber \\[-0.2cm] 
    & & \mbox{} + M^8(34\sh^7 + 179\sh^6\th + 356\sh^5\th^2 + 
          473\sh^4\th^3 + 473\sh^3\th^4 + 356\sh^2\th^5 + 
          179\sh\th^6 + 34\th^7) 
          \nonumber \\[-0.2cm] 
    & & \mbox{} - M^{10}(44\sh^6 + 193\sh^5\th + 346\sh^4\th^2 + 
          410\sh^3\th^3 + 346\sh^2\th^4 + 193\sh\th^5 + 44\th^6) 
          \nonumber \\[-0.2cm] 
    & & \mbox{} + M^{12}(34\sh^5 + 124\sh^4\th + 
          188\sh^3\th^2 + 188\sh^2\th^3 + 124\sh\th^4 + 34\th^5)
          \nonumber \\[-0.2cm] 
    & & \mbox{} - M^{14}(15\sh^4 + 43\sh^3\th + 52\sh^2\th^2 + 43\sh\th^3 + 
          15\th^4)
          \nonumber \\[-0.2cm] 
    & & \mbox{} + M^{16}(3\sh^3 + 6\sh^2\th + 6\sh\th^2 + 3\th^3) \Big\}
          \\
  b &=& 4\sh^{12} + 24\sh^{11}\th + 68\sh^{10}\th^2 + 124\sh^9\th^3 + 
          164\sh^8\th^4 + 176\sh^7\th^5 + 176\sh^6\th^6 
          \nonumber \\[-0.2cm]
    & & \mbox{} + 176\sh^5\th^7 + 164\sh^4\th^8 + 124\sh^3\th^9 + 
          68\sh^2\th^{10} + 24\sh\th^{11} + 4\th^{12}  
          \nonumber \\[-0.2cm]
    & & \mbox{} - M^2(20\sh^{11} + 104\sh^{10}\th + 250\sh^9\th^2 + 
          397\sh^8\th^3 + 481\sh^7\th^4 + 500\sh^6\th^5 
          \nonumber \\[-0.2cm]
    & & \mbox{} + 500\sh^5\th^6 + 481\sh^4\th^7 + 397\sh^3\th^8 + 
          250\sh^2\th^9 + 104\sh\th^{10} + 20\th^{11})
          \nonumber \\[-0.2cm]
    & & \mbox{} + M^4(40\sh^{10} + 166\sh^9\th + 278\sh^8\th^2 + 
          285\sh^7\th^3 + 206\sh^6\th^4 + 146\sh^5\th^5 + 
          206\sh^4\th^6 
          \nonumber \\[-0.2cm]
    & & \mbox{} + 285\sh^3\th^7 + 278\sh^2\th^8 + 
          166\sh\th^9 + 40\th^{10})
          \nonumber \\[-0.2cm]
    & & \mbox{} + M^6(-40\sh^9 - 97\sh^8\th + 53\sh^7\th^2 + 373\sh^6\th^3 + 
          647\sh^5\th^4 + 647\sh^4\th^5 + 373\sh^3\th^6 + 
          53\sh^2\th^7 
          \nonumber \\[-0.2cm]
    & & \mbox{} - 97\sh\th^8 - 40\th^9)
          \nonumber \\[-0.2cm]
    & & \mbox{} + M^8(20\sh^8 - 33\sh^7\th - 368\sh^6\th^2 - 751\sh^5\th^3 - 
          920\sh^4\th^4 - 751\sh^3\th^5 - 368\sh^2\th^6 
          \nonumber \\[-0.2cm]
    & & \mbox{} - 33\sh\th^7 + 20\th^8)
          \nonumber \\[-0.2cm]
    & & \mbox{} + M^{10}(-4\sh^7 + 77\sh^6\th + 323\sh^5\th^2 + 
          492\sh^4\th^3 + 492\sh^3\th^4 + 323\sh^2\th^5 + 
          77\sh\th^6 - 4\th^7)
          \nonumber \\[-0.2cm]
    & & \mbox{} - M^{12}(41\sh^5\th + 120\sh^4\th^2 + 142\sh^3\th^3 + 
          120\sh^2\th^4 + 41\sh\th^5)
          \nonumber \\[-0.2cm]
    & & \mbox{} + M^{14}(8\sh^4\th + 16\sh^3\th^2 + 16\sh^2\th^3 + 8\sh\th^4)
          \\
  c &=& -4\sh^{10}\th^2 - 20\sh^9\th^3 - 40\sh^8\th^4 - 40\sh^7\th^5 + 
           8\sh^6\th^6 + 80\sh^5\th^7 + 128\sh^4\th^8 + 116\sh^3\th^9
          \nonumber \\[-0.2cm]
    & & \mbox{} + 68\sh^2\th^{10} + 24\sh\th^{11} + 4\th^{12} 
          \nonumber \\[-0.2cm]
    & & \mbox{} + M^2(20\sh^9\th^2 + 56\sh^8\th^3 + 24\sh^7\th^4 - 
          147\sh^6\th^5 - 409\sh^5\th^6 - 599\sh^4\th^7 - 571\sh^3\th^8
          \nonumber \\[-0.2cm]
    & & \mbox{} - 370\sh^2\th^9 - 148\sh\th^{10} - 28\th^{11})
          \nonumber \\[-0.2cm]
    & & \mbox{} +  M^4(4\sh^{10} + 20\sh^9\th - 16\sh^8\th^2 - 48\sh^7\th^3 + 
          150\sh^6\th^4 + 611\sh^5\th^5 + 1060\sh^4\th^6 
          \nonumber \\[-0.2cm]
    & & \mbox{} + 1155\sh^3\th^7 + 854\sh^2\th^8 + 394\sh\th^9 + 84\th^{10})
          \nonumber \\[-0.2cm]
    & & \mbox{} - M^6(20\sh^9 + 88\sh^8\th + 48\sh^7\th^2 + 12\sh^6\th^3 + 
          318\sh^5\th^4 + 863\sh^4\th^5 + 1195\sh^3\th^6 
          \nonumber \\[-0.2cm]
    & & \mbox{} + 1061\sh^2\th^7 + 583\sh\th^8 + 140\th^9)
          \nonumber \\[-0.2cm]
    & & \mbox{} + M^8(40\sh^8 + 152\sh^7\th + 94\sh^6\th^2 + 38\sh^5\th^3 + 
          290\sh^4\th^4 + 631\sh^3\th^5 
          \nonumber \\[-0.2cm]
    & & \mbox{} + 738\sh^2\th^6 + 513\sh\th^7 + 140\th^8)
          \nonumber \\[-0.2cm]
    & & \mbox{} - M^{10}(40\sh^7 + 129\sh^6\th + 53\sh^5\th^2 
          + 7\sh^4\th^3 + 129\sh^3\th^4 + 264\sh^2\th^5 + 266\sh\th^6 
          + 84\th^7)
          \nonumber \\[-0.2cm]
    & & \mbox{} + M^{12}(20\sh^6 + 55\sh^5\th + 2\sh^4\th^2 - 15\sh^3\th^3 + 
          30\sh^2\th^4 + 76\sh\th^5 + 28\th^6)
          \nonumber \\[-0.2cm]
    & & \mbox{} + M^{14}(-4\sh^5 - 11\sh^4\th + 7\sh^3\th^2 + 7\sh^2\th^3 - 
          11\sh\th^4 - 4\th^5)
          \nonumber \\[-0.2cm]
    & & \mbox{} + M^{16}(\sh^3\th - 2\sh^2\th^2 + \sh\th^3)
          \\
  d &=& -2\sh^{10}\th^2 - 6\sh^9\th^3 - 2\sh^8\th^4 + 28\sh^7\th^5 + 
          88\sh^6\th^6 + 148\sh^5\th^7 + 166\sh^4\th^8 + 
          130\sh^3\th^9 
          \nonumber \\[-0.2cm]
    & & \mbox{} + 70\sh^2\th^{10} + 24\sh\th^{11} + 4\th^{12} 
          \nonumber \\[-0.2cm]
    & & \mbox{} + M^2 (4\sh^{11} + 22\sh^{10}\th + 70\sh^9\th^2 + 
          115\sh^8\th^3 + 71\sh^7\th^4 - 119\sh^6\th^5 - 381\sh^5\th^6  
          \nonumber \\[-0.2cm]
    & & \mbox{} - 552\sh^4\th^7 - 512\sh^3\th^8 - 320\sh^2\th^9 - 
          126\sh\th^{10} - 24\th^{11})
          \nonumber \\[-0.2cm]
    & & \mbox{} + M^4(-20\sh^{10} - 104\sh^9\th - 296\sh^8\th^2 - 
          459\sh^7\th^3 - 352\sh^6\th^4 + 73\sh^5\th^5 + 
          558\sh^4\th^6 
          \nonumber \\[-0.2cm]
    & & \mbox{} + 744\sh^3\th^7 + 574\sh^2\th^8 + 
          270\sh\th^9 + 60\th^{10})
          \nonumber \\[-0.2cm]
    & & \mbox{} + M^6(40\sh^9 + 199\sh^8\th + 533\sh^7\th^2 + 
          778\sh^6\th^3 + 596\sh^5\th^4 + 51\sh^4\th^5 - 405\sh^3\th^6 
          \nonumber \\[-0.2cm]
    & & \mbox{} - 480\sh^2\th^7 - 296\sh\th^8 - 80\th^9) 
          \nonumber \\[-0.2cm]
    & & \mbox{} + M^8(-40\sh^8 - 197\sh^7\th - 506\sh^6\th^2 - 
          672\sh^5\th^3 - 460\sh^4\th^4 - 79\sh^3\th^5 + 138\sh^2\th^6 
          \nonumber \\[-0.2cm]
    & & \mbox{} + 164\sh\th^7 + 60\th^8)
          \nonumber \\[-0.2cm]
    & & \mbox{} + M^{10}(20\sh^7 + 107\sh^6\th + 267\sh^5\th^2 + 
         307\sh^4\th^3 + 185\sh^3\th^4 + 56\sh^2\th^5 - 
         30\sh\th^6 - 24\th^7)
          \nonumber \\[-0.2cm]
    & & \mbox{} + M^{12}(-4\sh^6 - 31\sh^5\th - 74\sh^4\th^2 
         - 71\sh^3\th^3 - 46\sh^2\th^4 - 10\sh\th^5 + 4\th^6)
          \nonumber \\[-0.2cm]
    & & \mbox{} + M^{14}(4\sh^4\th + 8\sh^3\th^2 + 8\sh^2\th^3 + 4\sh\th^4)
\eeqa

\noindent $g + q
\rightarrow c\bar c \left[ ^3S_1^{(8)} \right] + q$:\\[-1.2cm]

\beqa
  F &=&  \frac{(4\pi\alpha_s)^3 \langle {\cal O}_8^{J/\psi}(^3\!S_1) 
           \rangle}{216 M^3 \sh\uh (\th-M^2)^2} 
           \left\{4 (\th-M^2)^2-9\sh\uh\right\} \\
  a &=& \sh^2+\uh^2+2 M^2\th \\
  b &=& 4 \\
  c &=& 8 \\
  d &=& 4
\eeqa

\noindent $g + q
\rightarrow c\bar c \left[ ^1S_0^{(8)} \right] + q$:\\[-1.2cm]

\beqa
  Fa &=& \frac{5 (4\pi\alpha_s)^3
           \langle {\cal O}_8^{J/\psi}(^1S_0) \rangle}{
           216 M \th (\th-M^2)^2} \left\{\sh^2+\uh^2\right\} \\
  b  &=& c = d = 0 
\eeqa

\noindent $g + q
\rightarrow c\bar c \left[ ^3P_J^{(8)} \right] + q$:\\[-1.2cm]

\beqa
  F &=& \frac{5 (4\pi\alpha_s)^3 \langle {\cal O}_8^{J/\psi}(^3\!P_0) 
          \rangle}{18 M^3\th^2 (\th-M^2)^3} \\
  a &=& \th (\th-M^2) (\sh^2+\uh^2+2 M^2\th+2 M^4) \\
  b &=& -8 (\sh^2+\sh\th+M^2\th) \\
  c &=& 8 (\th^2-M^4) \\
  d &=& 4 (\th^2-2 M^2\sh-M^2\th)
\eeqa

\noindent $q + \bar{q}
\rightarrow c\bar c \left[ ^3S_1^{(8)} \right] + g$:\\[-1.2cm]

\beqa
  F &=&  -\frac{(4\pi\alpha_s)^3 \langle {\cal O}_8^{J/\psi}(^3\!S_1) 
           \rangle}{81 M^3 \th\uh (\sh-M^2)^2} \,\left\{
           4 (\sh-M^2)^2-9\th\uh\right\} \\
  a &=& \th^2+\uh^2+2 M^2\sh \\
  b &=& 4 \\
  c &=& 4 \\
  d &=& 0
\eeqa

\noindent $q + \bar{q}
\rightarrow c\bar c \left[ ^1S_0^{(8)} \right] + g$:\\[-1.2cm]

\beqa
  Fa &=& -\frac{5 (4\pi\alpha_s)^3
           \langle {\cal O}_8^{J/\psi}(^1S_0) \rangle}{
           81 M \sh (\sh-M^2)^2} \left\{\th^2+\uh^2\right\} \\
  b  &=& c = d = 0 
\eeqa

\noindent $q + \bar{q}
\rightarrow c\bar c \left[ ^3P_J^{(8)} \right] + g$:\\[-1.2cm]

\beqa
  F &=& \frac{20 (4\pi\alpha_s)^3 \langle {\cal O}_8^{J/\psi}(^3\!P_0) 
          \rangle}{27 M^3\sh^2 (\sh-M^2)^3}\\
  a &=& -\sh (\sh-M^2) (\th^2+\uh^2+2 M^2\sh+2 M^4) \\
  b &=& 8 (\th^2+\sh\th+M^2\sh) \\
  c &=& 8 (\th^2+\sh\th-2 M^2\th+M^4) \\
  d &=& 4 (\sh^2+2\sh\th+2\th^2+M^2\sh-2 M^2\th)
\eeqa

\end{document}